\documentclass[aps,pra,reprint,groupedaddress]{revtex4-2}

\usepackage{amsmath,amssymb,amsthm,bm,color,mathrsfs,mathtools,subcaption,xcolor} 
\usepackage[setpagesize=false]{hyperref}
\usepackage[pdftex]{graphicx}
\newcommand{\T}{\mathrm{T}}
\newcommand{\f}[2]{\frac{#1}{#2}}

\newcommand{\lt}{\left}
\newcommand{\rt}{\right}
\newcommand{\pd}{\partial}
\newcommand{\ket}[1]{|{#1}\rangle}
\newcommand{\bra}[1]{\langle{#1}|}
\newcommand{\braket}[1]{\langle{#1}\rangle}
\newcommand{\mcal}[1]{\mathcal{#1}}
\DeclareMathOperator\tr{Tr}

\DeclareMathOperator*\argmin{arg\,min}
\DeclareMathOperator*\argmax{arg\,max}

\begin{document}


\title{Adaptive quantum state estimation for two optical point sources}



\author{Masataka Kimizu}
\email{kimizu@sigmath.es.osaka-u.ac.jp}
\affiliation{Graduate School of Engineering Science, Osaka University, Toyonaka, Osaka 560-0043, Japan}

\author{Fuyuhiko Tanaka}
\email{ftanaka.celas@osaka-u.ac.jp}
\affiliation{Center for Education in Liberal Arts and Sciences, Osaka University, Toyonaka, Osaka 560-0043, Japan}
\affiliation{Center for Quantum Information and Quantum Biology, Osaka University, Toyonaka, Osaka 560-0043, Japan}

\author{Akio Fujiwara}
\email{fujiwara@math.sci.osaka-u.ac.jp}
\affiliation{Department of Mathematics, Osaka University, Toyonaka, Osaka 560-0043, Japan}
\affiliation{Center for Quantum Information and Quantum Biology, Osaka University, Toyonaka, Osaka 560-0043, Japan}


\date{\today}

\begin{abstract} 
In classical optics, there is a well-known resolution limit, called Rayleigh's curse, in the separation of two incoherent optical sources in close proximity.
Recently, Tsang {\it et al.} revealed that this difficulty may be circumvented in the framework of quantum theory. Following their work, various estimation methods have been proposed to overcome Rayleigh's curse, but none of them enables us to estimate the positions of two point sources simultaneously based on single-photon measurements with high accuracy. In this study, we propose a method to {\it simultaneously} estimate the positions of two point sources with the highest accuracy using adaptive quantum state estimation scheme. 
\end{abstract}


\maketitle

\section{Introduction}

Discriminating two optical point sources is an important subject in optics that is expected to be applied to astronomical observations and biological imaging. However, the conventional method has a drawback called Rayleigh's curse \cite{tsang16}, which makes it difficult to discriminate two point sources when they are close to each other. 
This problem can be translated as that of estimating the centroid and the separation of two point sources, and Rayleigh's curse represents the difficulty in estimating the separation when two point sources are close to each other. Recently, Tsang {\it et al.} \cite{tsang16} investigated this problem in the framework of quantum theory and showed that there is a possibility of estimating the separation of two point sources in close proximity with the same accuracy as when they are far apart. 
Moreover, they devised a measurement scheme called the spatial mode demultiplexing (SPADE) that achieves this accuracy when the centroid of two point sources is known in advance.

The scheme SPADE allows us to accurately estimate the separation, but it requires prior knowledge of the centroid. 
Accordingly, a two-step procedure was proposed by Grace {\it et al.} \cite{grace20} in which the centroid was to be  estimated first.
Meanwhile, Parniak {\it et al.} \cite{parniak18} and Bao {\it et al.} \cite{bao21} investigated simultaneous estimation of the centroid and the separation, but they did not take account of the optimality of the measurement.

The optimal measurement for multiple parameters can be obtained from the simultaneous spectral decompositions of the symmetric logarithmic derivatives (SLD) if they commute.
Unfortunately, the SLDs of the centroid and the separation of two point sources do not in general commute \cite{shi23}.
In such cases, it is customary to search for a measurement that minimizes the weighted trace of the covariance matrix (or that of the inverse Fisher information matrix) \cite{holevo11,nagaoka89,fujiwara06,yamagata11}. Once the optimal measurement is obtained, the parameters can be estimated simultaneously with high accuracy using an estimation scheme called adaptive quantum state estimation (AQSE), which was proposed by Nagaoka \cite{nagaoka89} and theoretically justified by Fujiwara \cite{fujiwara06}. Since the optimal measurement generally depends on the true values of the parameters, AQSE updates the measurement sequentially.

In this study, we propose a method to {\it simultaneously} estimate the centroid and the separation of two point sources using AQSE. In particular, the measurement we use is the {\it optimal} one for estimating both the centroid and the separation, and the weighted trace of the sample covariance matrix is asymptotically the smallest in theory. Through numerical experiments, we confirm that the proposed method works effectively if the number of steps in AQSE is sufficiently large. 

The paper is organized as follows. In Sec.~\ref{sec:settings}, we describe the mathematical formulation of our estimation problem. 
In Sec.~\ref{sec:review}, we briefly summarize related works such as direct imaging and SPADE. 
In Sec.~\ref{sec:AQSE}, we first introduce an AQSE scheme for two optical point sources using numerically obtained optimal measurements, 
and then carry out numerical simulations of AQSE to demonstrate that the centroid and the separation can in principle be estimated simultaneously with the best accuracy in the asymptotic limit.
We also find a significant reduction in the rate of convergence of estimates as the separation of two point sources gets closer to zero; this phenomenon may correspond to Rayleigh's curse.
Finally, we summarize the paper in Sec.~\ref{sec:summary}.

\section{Problem setting} \label{sec:settings}

In this section, we present the mathematical formulation of the problem we consider mainly based on Tsang {\it et al.} \cite{tsang16}.

\subsection{Mathematical formulation of our problem}

The light emitted from two point sources is assumed to be quasi-monochromatic and of equal brightness, and the image plane is assumed to be one-dimensional. Let $\epsilon\ll1$ be the average number of photons observed at each temporal mode. The density operator in the image plane at each temporal mode is
\begin{equation}
\rho = (1-\epsilon)\rho_0+\epsilon\rho_1+O(\epsilon^2)
\end{equation}
where $\rho_0$ is the zero-photon state and $\rho_1$ is the one-photon state. Since two or more photons are almost never observed simultaneously in a single measurement when $\epsilon\ll1$, we shall focus our attention only on the one-photon state $\rho_1$. 

We write $L^2(\mathbb{R})$ for the set of square integrable real-valued functions on $\mathbb{R}$. Let $\ket{\psi_1},\ket{\psi_2}\in L^2(\mathbb{R})$ denote the states in the image plane of a single-photon emitted from each point source. Then, $\rho_1$ can be written as
\begin{equation}
\rho_1 = \f12\lt(\ket{\psi_1}\bra{\psi_1}+\ket{\psi_2}\bra{\psi_2}\rt).
\end{equation}
This equation is in fact an approximation, but we will treat it as accurate. We assume that $\ket{\psi_1}$ and $\ket{\psi_2}$ are expressed as
\begin{equation}
\ket{\psi_j} = \int_{-\infty}^\infty dx\,\psi(x-x_j)\ket{x},\qquad j=1,2
\end{equation}
with $\psi(x)$ being the point-spread function and $x_j$ the coordinates of the $j$th point source satisfying $x_1<x_2$. Here, $\ket{x}$ represents the ideal state in which the photon is localized exactly at position $x$.
In this paper, we assume that the point-spread function $\psi(x)$ is Gaussian:
\begin{equation} \label{eq:G-psf}
\psi(x) = \f{1}{\lt(2\pi\sigma^2\rt)^{\f14}}\exp\lt(-\f{x^2}{4\sigma^2}\rt)
\end{equation}
where $\sigma$ is a positive constant determined by the wavelength of the light and the properties of the lens.

Our problem is to estimate the true values of the coordinates $x_1$ and $x_2$, or equivalently, the transformed parameters
\begin{equation}
\theta^1 = \f{x_1+x_2}{2},\qquad \theta^2 = x_2-x_1
\end{equation}
simultaneously. In what follows, we call $\theta^1$ the centroid and $\theta^2$ the separation, and denote $\rho_1$ as $\rho_\theta$, where $\theta=(\theta^1,\theta^2)\in\Theta=\mathbb{R}\times\mathbb{R}_{>0}$.

\subsection{Cram\'er--Rao bound and quantum Fisher information matrix}

In order to estimate the true values of the parameters, we apply a measurement $M=\lt\{M(\omega)\mid\omega\in\Omega\rt\}$ represented by a POVM to a one-photon state $\rho_\theta$, where $\Omega$ is the set of measurement outcomes. Here, the measurement can be chosen arbitrarily, but once it is fixed, $p_\theta(\omega;M)=\tr\rho_\theta M(\omega)$ gives the probability distribution of the outcomes. This allows us to consider the Cram\'er--Rao inequality 
\begin{equation} \label{eq:cCR}
V_\theta[M,\hat\theta] \ge J_\theta(M)^{-1}
\end{equation}
which gives a lower bound on the estimation error for any (locally) unbiased estimator $\hat\theta$, where $V_\theta[M,\hat\theta]$ is the covariance matrix and $J_\theta(M)$ is the (classical) Fisher information matrix of the parametric model $p_\theta({}\cdot{}; M)$. In particular, given a measurement $M$, there is a locally unbiased estimator $\hat\theta$ that achieves the lower bound in \eqref{eq:cCR} \cite{holevo11}. The asymptotic lower bound for the precision of the maximum likelihood estimator (MLE), which is not in general locally unbiased, is also characterized by $J_\theta(M)^{-1}$. 

In quantum estimation theory, we often consider the quantum Fisher information matrix instead of the Fisher information matrix to evaluate the error bound. The quantum Fisher information matrix $K_\theta$ is defined using the Hermitian operator $L_j$ satisfying the following equation
\begin{equation}
\f{\pd\rho_\theta}{\pd\theta^j} = \f12\lt(\rho_\theta L_j+L_j\rho_\theta\rt).
\end{equation}
The operator $L_j$ is called the symmetric logarithmic derivative (SLD) in the direction $\theta^j$. The $(j,k)$th component of the matrix $K_\theta$ is defined as
\begin{equation}
\lt(K_\theta\rt)_{jk} = \f12\tr\rho_\theta\lt(L_jL_k+L_kL_j\rt).
\end{equation}

It is known that the Fisher information matrix $J_\theta(M)$ for a given measurement $M$ is bounded from above by the quantum Fisher information matrix $K_\theta$, in that,
\begin{equation} \label{eq:QCR}
 J_\theta(M) \le K_\theta.
\end{equation}
If there exists a measurement $M$ that achieves the upper bound in \eqref{eq:QCR}, it is the optimal measurement. Such a measurement exists when the SLDs $\{L_j\}$ commute, but it is not always the case.

In our model, the quantum Fisher information matrix of $\rho_\theta$ with respect to the parameter $\theta^1$ and $\theta^2$ is written as
\begin{equation} \label{eq:QFIM}
K_\theta = \lt(\begin{array}{cc} \f{1}{\sigma^2}-\f{(\theta^2)^2}{4\sigma^4}\exp\lt(-\f{(\theta^2)^2}{4\sigma^2}\rt) & 0 \\ 0 & \f{1}{4\sigma^2} \end{array}\rt).
\end{equation}
See \cite{tsang16} for derivation. It is important to realize that, unless $\theta^2=2\sigma$, the two SLDs do not commute, and there is no measurement that achieves the upper bound in \eqref{eq:QCR}, (cf., Fig.~\ref{fig:opt_v}). Thus, we have to find an optimal measurement by another approach. We shall discuss this issue again in Sec.~\ref{subsec:opt_meas}.

\section{Brief review of previous studies} \label{sec:review}

In this section, we briefly review the conventionally considered measurement and those proposed in related studies.

\subsection{Direct imaging}

Direct imaging is a simple method of measuring the position of a photon. The probability distribution of direct imaging is
\begin{align}
p_\theta&(x;M_{\rm direct}) = \tr\rho_\theta\ket{x}\bra{x} = \f12\lt(|\braket{x|\psi_1}|^2+|\braket{x|\psi_2}|^2\rt) \notag \\
&= \f12\lt(\lt|\psi(x-x_1)\rt|^2+\lt|\psi(x-x_2)\rt|^2\rt) \notag \\
&= \f12\lt(\lt|\psi(x-(\theta^1-\theta^2/2))\rt|^2+\lt|\psi(x-(\theta^1+\theta^2/2))\rt|^2\rt) 
\end{align}
which is a mixture of Gaussian distributions centered at $\theta^1$ and shifted by $\pm\theta^2/2$.

In this measurement, when $\theta^2\ll\sigma$, it is easy to estimate $\theta^1$ because the Fisher information for $\theta^1$ is almost equal to the quantum Fisher information, but it is difficult to estimate $\theta^2$ because the Fisher information for $\theta^2$ converges to zero in the limit of $\theta^2\downarrow0$ \cite{tsang16}. This fact is a variant of Rayleigh's curse in view of statistical estimation.

\subsection{HG SPADE}

Notwithstanding the above-mentioned fact, there is a room for improving the precision of the estimation of $\theta^2$ by means of a different type of measurement, since the quantum Fisher information for $\theta^2$ is a positive constant $1/(4\sigma^2)$ as seen from \eqref{eq:QFIM}.

HG SPADE is a measurement proposed by Tsang {\it et al.} \cite{tsang16} to improve the accuracy of the estimation of $\theta^2$. In HG SPADE, assuming that the estimate $\hat\theta^1$ of the centroid is obtained a priori, the measurement is performed by the POVM $M_{\rm HG}=\lt\{\ket{\phi_q}\bra{\phi_q}\mid q=0,1,\hdots\rt\}$ where
\begin{align}
\ket{\phi_q} &= \int_{-\infty}^\infty dx\phi_q(x-\hat\theta^1)\ket{x},\qquad q=0,1,\hdots, \\
\phi_q(x) &= \f{1}{\lt(2\pi\sigma^2\rt)^{\f14}}\f{1}{\sqrt{2^qq!}}H_q\lt(\f{x}{\sqrt{2}\sigma}\rt)\exp\lt(-\f{x^2}{4\sigma^2}\rt),
\end{align}
and $H_q(x)$ is the Hermite polynomial. That is, we do not measure the position of the photon, but rather which spatial mode $\ket{\phi_q}$ the photon is in. HG stands for Hermite-Gaussian and SPADE for spatial-mode demultiplexing. The probability $p_\theta(q;M_{\rm HG})$ of obtaining the measurement outcome $q$ is
\begin{align}
p_\theta(q;M_{\rm HG}) &= \tr\rho_\theta\ket{\phi_q}\bra{\phi_q} \notag \\
&= \f12\lt(|\braket{\phi_q|\psi_1}|^2+|\braket{\phi_q|\psi_2}|^2\rt) \notag \\
&= \f12\lt(\exp(-Q_1)\f{Q_1^q}{q!}+\exp(-Q_2)\f{Q_2^q}{q!}\rt),
\end{align}
where
\begin{align}
Q_1 &= \f{1}{4\sigma^2}\lt(\hat\theta^1-\theta^1+\f{\theta^2}{2}\rt)^2,\\
Q_2 &= \f{1}{4\sigma^2}\lt(\hat\theta^1-\theta^1-\f{\theta^2}{2}\rt)^2.
\end{align}
Namely, $p_\theta(q;M_{\rm HG})$ is a mixture of two Poisson distributions with parameters $Q_1$ and $Q_2$, respectively.

Assume now that the value of the centroid is known exactly, $\hat\theta^1=\theta^1$. Then
\begin{align}
p_\theta(q;M_{\rm HG}) &= |\braket{\phi_q|\psi_1}|^2 = |\braket{\phi_q|\psi_2}|^2 = \exp(-Q)\f{Q^q}{q!},
\end{align}
where
\begin{align}
Q &= \f{(\theta^2)^2}{16\sigma^2},
\end{align}
and the Fisher information of this probability distribution for $\theta^2$ is
\begin{align}
J_\theta(M_{\rm HG})_{22} &= \sum_{q=0}^\infty p_\theta(q;M_{\rm HG})\lt\{\f{\pd}{\pd\theta^2}\log p_\theta(q;M_{\rm HG})\rt\}^2 \notag \\
&= \f{1}{4\sigma^2}.
\end{align}
This is identical to the $(2,2)$th entry of the quantum Fisher information matrix \eqref{eq:QFIM}. In other words, if the true value of the centroid $\theta^1$ is known exactly, HG SPADE is the best measurement for estimating the separation $\theta^2$.

However, it is also pointed out in \cite{tsang16} that, if the estimate of the centroid $\hat\theta^1$ deviates even slightly from the true value, the Fisher information for $\theta^2$ falls to zero in the limit of $\theta^2\downarrow0$.

\subsection{Some other studies}

Since HG SPADE requires accurate knowledge of the centroid $\theta^1$, a two-step procedure was proposed by Grace {\it et al.}  \cite{grace20} in which $\theta^1$ was first estimated by direct imaging and then $\theta^2$ was estimated by SPADE. 
Meanwhile, simultaneous estimation of $\theta^1$ and $\theta^2$ was studied by Parniak {\it et al.} \cite{parniak18} and Bao {\it et al.} \cite{bao21}. Parniak {\it et al.} \cite{parniak18} used quantum correlation to measure two photons together and did not investigate simultaneous estimation with single-photon measurements without quantum correlations. Their measurement is physically feasible, but is not necessarily optimal. Bao {\it et al.} \cite{bao21}, on the other hand, took a Bayesian approach to simultaneous estimation, but they also did not take account of the optimality of the measurement.

\section{Adaptive parameter estimation} \label{sec:AQSE}

Adaptive quantum state estimation (AQSE), proposed by Nagaoka \cite{nagaoka89} and theoretically justified by Fujiwara \cite{fujiwara06}, is an efficient estimation scheme for unknown parameters of a given quantum statistical model. In this section, we first briefly describe this estimation scheme, and then apply it for the problem of estimating the positions of two point sources simultaneously.

\subsection{Protocol} \label{subsec:protocol}

Given a quantum statistical model $\lt\{\rho_\theta\mid\theta\in\Theta\subset\mathbb{R}^d\rt\}$, let $\theta_*$ be the true value of the parameter and write $M({}\cdot{};\theta_*)$ for its optimal measurement, taking account of the fact that, in general, the optimal measurement depends on the unknown true value of the parameter. In order to circumvent this difficulty, one may invoke an AQSE protocol, which runs as follows: choose the initial estimate $\hat\theta_0\in\Theta$ arbitrarily and repeat the following (i) and (ii) for step $n=1$, 2, \ldots.
\begin{enumerate}
\renewcommand{\labelenumi}{(\roman{enumi})}
\item Apply the measurement $M({}\cdot{};\hat\theta_{n-1})$, which is optimal at the previous estimate $\hat\theta_{n-1}$, to yield the $n$th outcome $\omega_n$.
\item Obtain the next estimate $\hat\theta_n$ from the data $(\omega_1,\hdots,\omega_n)$ by the maximum likelihood method, i.e.,
\begin{equation}
\hat\theta_n = \argmax_{\theta\in\Theta}\prod_{i=1}^n\tr\rho_\theta M(\omega_i;\hat\theta_{i-1}).
\end{equation}
\end{enumerate}

It was shown in \cite{fujiwara06} that, under some regularity conditions, $\hat\theta_n$ enjoys the strong consistency:
\begin{equation} \label{eq:sc}
\hat\theta_n \longrightarrow \theta_* \quad\text{with prob. 1}
\end{equation}
and the asymptotic efficiency:
\begin{equation} \label{eq:ae}
\sqrt{n}\bigl(\hat\theta_n-\theta_*\bigr) \longrightarrow N\lt(0,J_{\theta_*}(M({}\cdot{};\theta_*))^{-1}\rt) \quad\text{in dist.}
\end{equation}
In actual experiments, $n$ cannot be infinitely large and must be stopped at some point. However, if it is stopped at a sufficiently large $n$, the left-hand side of \eqref{eq:ae} approximately follows the distribution of the right-hand side, and a good estimation accuracy can be obtained.

\subsection{Optimal measurement in simultaneous estimation of \texorpdfstring{$\theta=(\theta^1,\theta^2)$}{θ=(θ1,θ2)}} \label{subsec:opt_meas}

Prior to applying AQSE, we need to obtain a list of optimal measurements $M({}\cdot{};\theta)$ for all $\theta\in\Theta$. Since the asymptotic fluctuation of the estimate $\hat\theta_n$ obtained by AQSE is characterized by the inverse of the Fisher information matrix $J_\theta(M)$ as in \eqref{eq:ae}, one may conceive that the optimal measurement would be the one that makes $J_\theta(M)^{-1}$ as small as possible.
But, in reality, one cannot minimize it since it is a matrix. 

One approach to finding the optimal measurement is to minimize the weighted trace of the inverse of the Fisher information matrix \cite{fujiwara06,nagaoka89,holevo11,yamagata11}:
\begin{equation} \label{eq:opt_m}
M({}\cdot{};\theta) = \argmin_{M:\text{ POVM}}\tr GJ_\theta(M)^{-1},
\end{equation}
given a positive definite matrix $G$, called the weight matrix, which may depend on the parameter $\theta$. 

In our problem, the underlying Hilbert space is $L^2(\mathbb{R})$, and thus $M({}\cdot{};\theta)$ in \eqref{eq:opt_m} must be obtained as a POVM on $L^2(\mathbb{R})$. However, as discussed in Shao {\it et al.} \cite{shao22}, $\rho_\theta,\,\f{\pd\rho_\theta}{\pd\theta^1}$ and $\f{\pd\rho_\theta}{\pd\theta^2}$ that appear in $J_\theta(M)$ have support on a $\theta$-dependent four-dimensional subspace $\mcal{V}_\theta$ of $L^2(\mathbb{R})$, and it is enough to obtain $M({}\cdot{};\theta)$ as a POVM on $\mcal{V}_\theta$; in fact, we need only add $I_{\mcal{V}_\theta^\perp}$ to obtain the POVM on $L^2(\mathbb{R})$. 

Unfortunately, the analytical solution for the minimization problem \eqref{eq:opt_m} is not known unless the underlying Hilbert space is two-dimensional \cite{yamagata11}. We therefore invoke numerical methods to find the optimal measurement $M({}\cdot{};\theta)$ for each $\theta$. Note that it is sufficient to consider 16-valued real rank-one measurements 
\footnote{It is theoretically known that according to Fujiwara \cite{fujiwara06}, 16 values are sufficient, but in fact, according to Yamagata \cite{yamagata_m}, 12 values are sufficient.},
and the minimization problem is reduced to an unconstrained nonlinear programming problem \cite{yamagata_m}; see Appendix for details. In what follows, we choose the weight matrix $G$ to be the quantum Fisher information matrix $K_\theta$. This choice is beneficial because the solution of the minimization problem \eqref{eq:opt_m} with this choice depends only on the state $\rho_\theta$ and is independent of the parametrization $\theta$.

\begin{figure}[tb]
  \captionsetup{singlelinecheck=false, justification=raggedright, font=small}
  \centering
  \includegraphics[width=85mm]{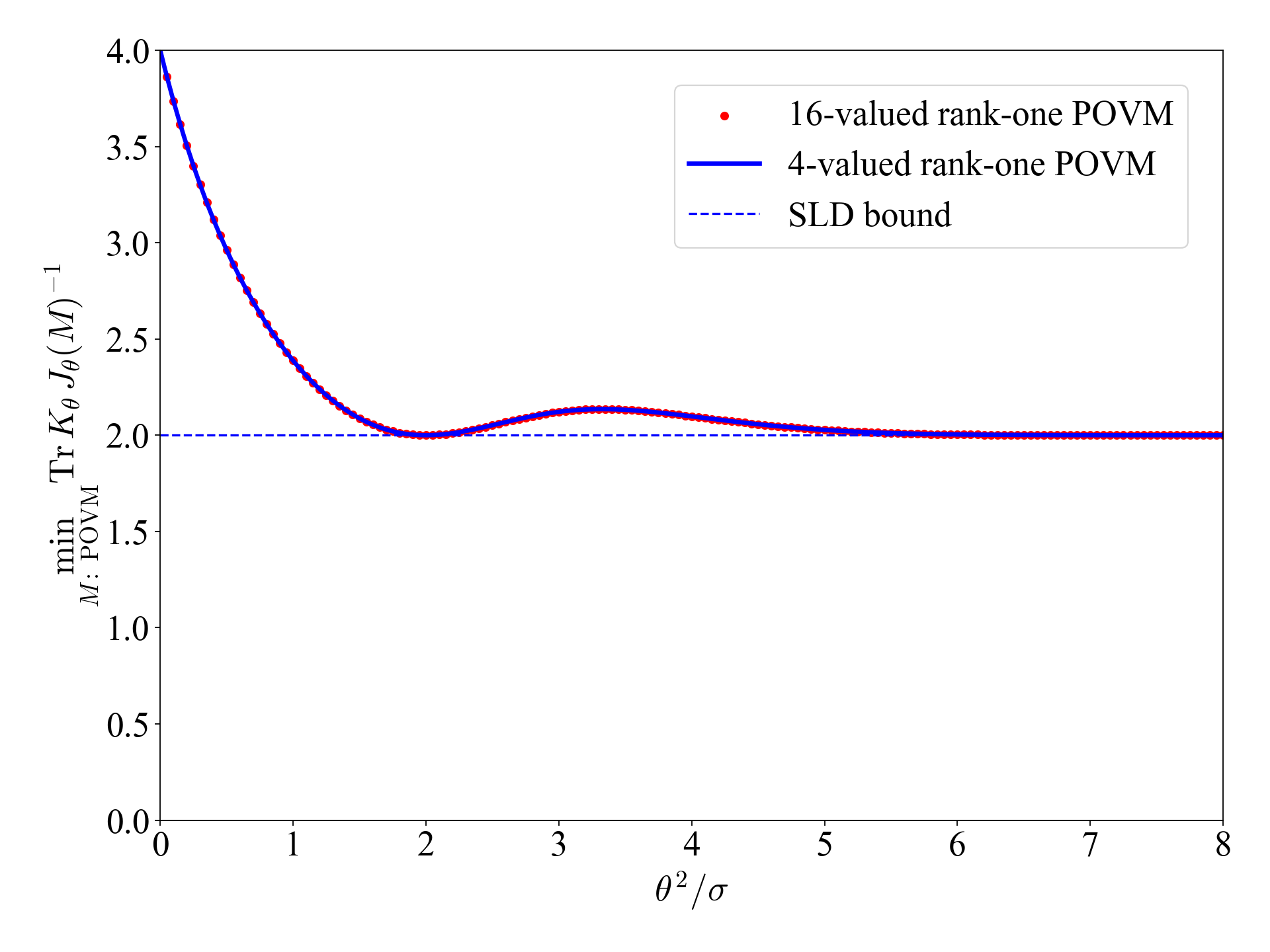}
  \caption{Minimum values of the weighted trace of the inverse Fisher information matrix. The blue dashed line is the SLD bound $\tr GK_\theta^{-1}=2$. The SLD bound is achieved when $\theta^2=2\sigma$ and nearly achieved when $\theta^2\gtrapprox6\sigma$.}
  \label{fig:opt_v}
\end{figure}

Fig.~\ref{fig:opt_v} demonstrates the solution for the minimization problem \eqref{eq:opt_m}. The horizontal axis is set to $\theta^2/\sigma$ because the minimum values depend only on $\theta^2$ due to the covariant nature of the model under parallel translation of the optical point sources. The red dots are the results of optimization with 16-valued rank-one measurements, and the blue dashed line shows the SLD bound $\tr GK_\theta^{-1}=2$. The figure shows that the SLD bound is achieved when $\theta^2=2\sigma$ and nearly achieved when $\theta^2\gtrapprox6\sigma$. This is because the SLDs for $\theta^1$ and $\theta^2$ can be taken to be commutative when $\theta^2=2\sigma$ \cite{shi23}, and the two SLDs are nearly commutative when $\theta^2$ is sufficiently large. Furthermore, the blue curve is the result of optimization with four-valued rank-one measurements, showing that the minimum value is achieved with four-valued measurement. In particular, this four-valued rank-one measurement is a projective measurement since $\mcal{V}_\theta$ is four-dimensional. Regarding the physical realization of projective measurements, Sajjad {\it et al.} \cite{sajjad21} have stated that ``any projective measurement on a quantum state of one photon in many (spatial) modes, which is the case for the quantum description of the state of a single temporal mode of collected light in our problem, is always realizable by a passive linear optical transformation followed by photon detection.''

\begin{figure}[tb]
  \captionsetup{singlelinecheck=false, justification=raggedright, font=small}
  \centering
  \includegraphics[width=85mm]{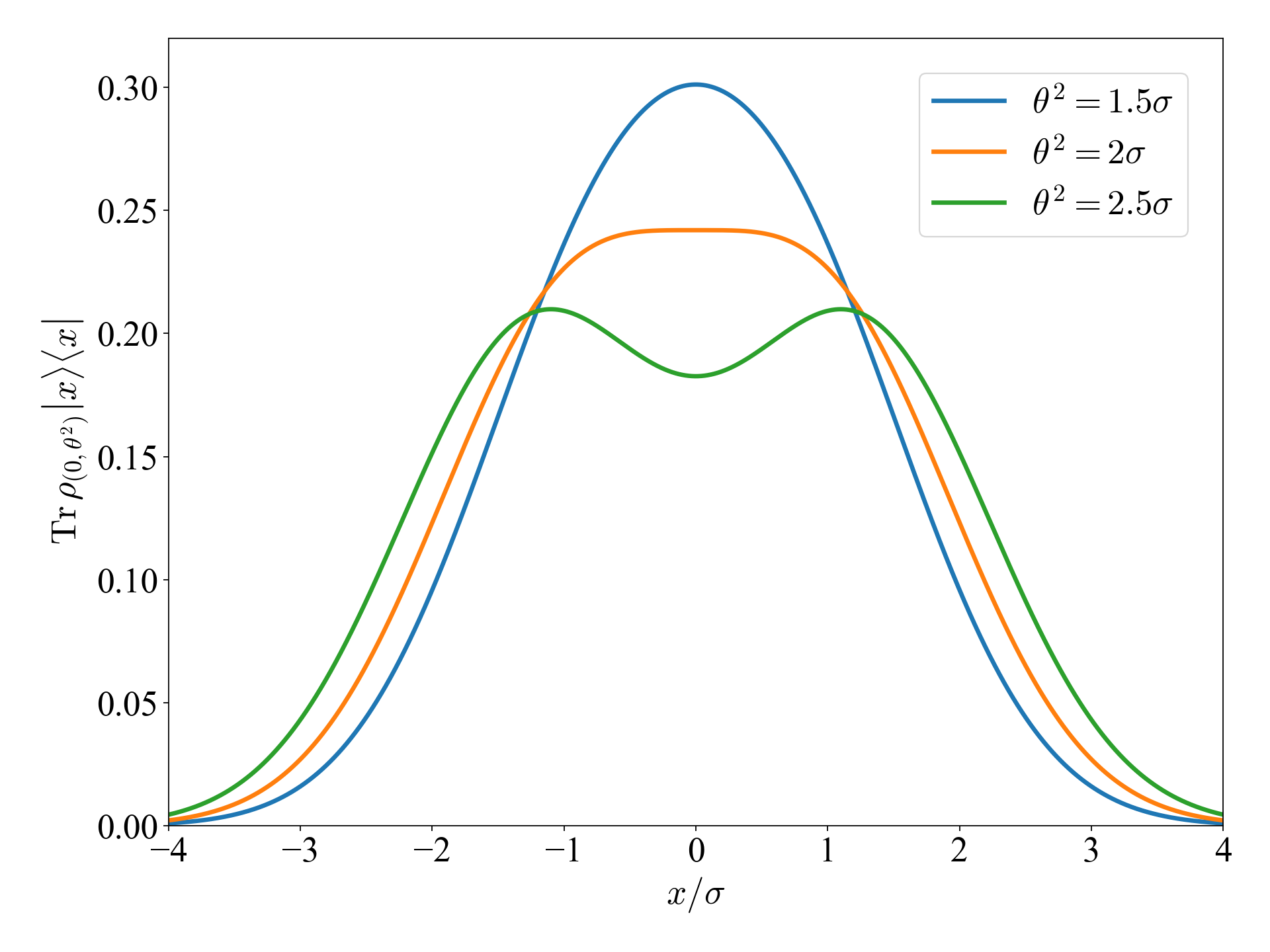}
  \caption{Probability distributions for direct imaging. The value $\theta^2=2\sigma$ is the boundary between unimodal and bimodal.}
  \label{fig:th2_2}
\end{figure}

Note that $\theta^2=2\sigma$ is the threshold at which the modality of the probability distributions for the direct imaging changes. In fact, as shown in Fig.~\ref{fig:th2_2}, the probability distribution for direct imaging on $\rho_{(0,\theta^2)}$ is unimodal when $0<\theta^2<2\sigma$, while it is bimodal when $\theta^2>2\sigma$.

\begin{figure}[tb]
  \captionsetup{singlelinecheck=false, justification=raggedright, font=small}
  \centering
  \includegraphics[width=85mm]{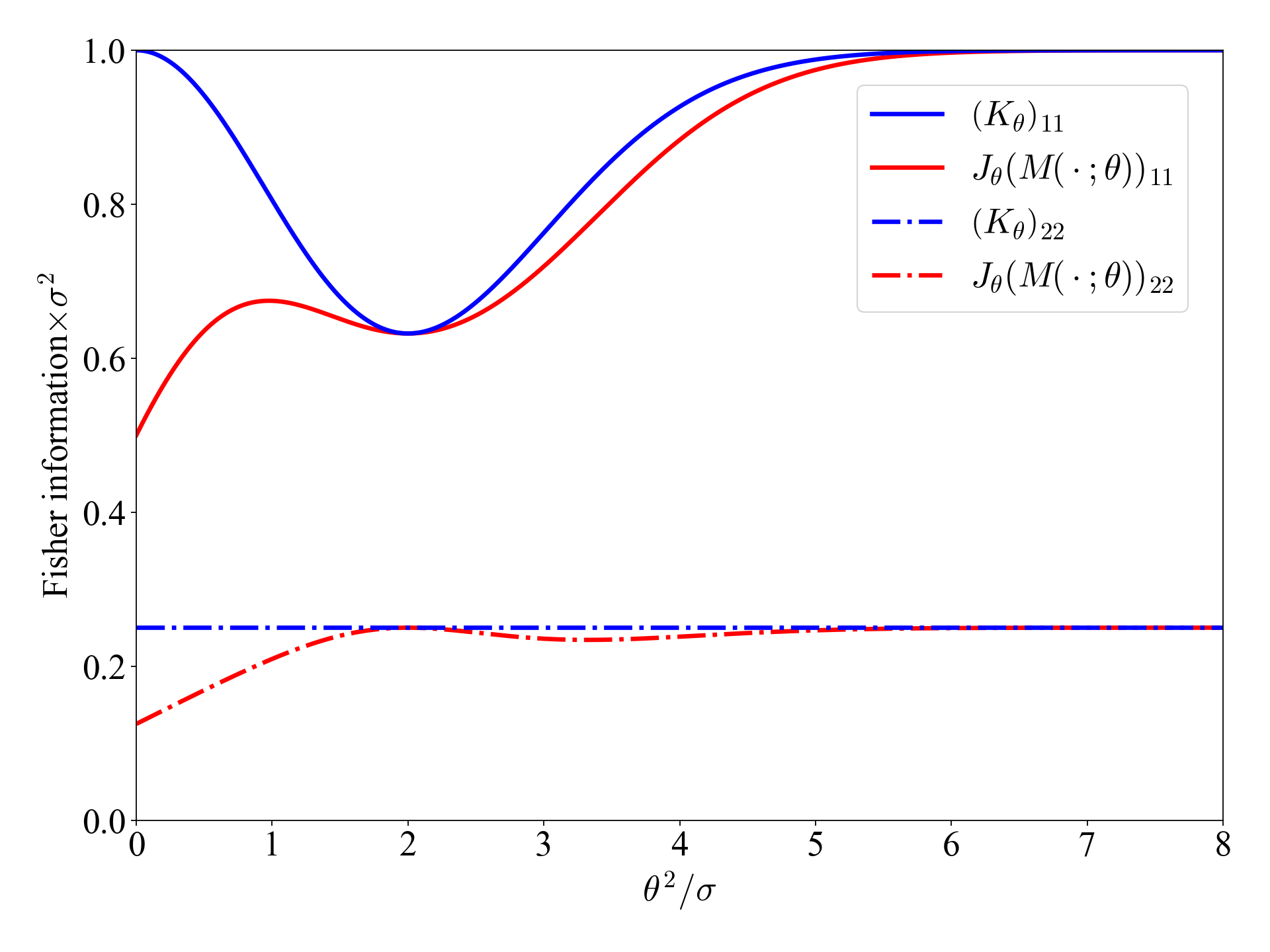}
  \caption{Comparison of the classical Fisher information under optimal measurements (red curves) with the quantum Fisher information (blue curves). 
Note that the classical Fisher information for $\theta^1$ and $\theta^2$ converges to positive values in the limit $\theta^2\downarrow0$.}
  \label{fig:opt_FI}
\end{figure}

\begin{figure*}[ht]
\begin{tabular}{cccc}
\begin{minipage}{0.24\hsize}
  \centering
  \includegraphics[width=43mm]{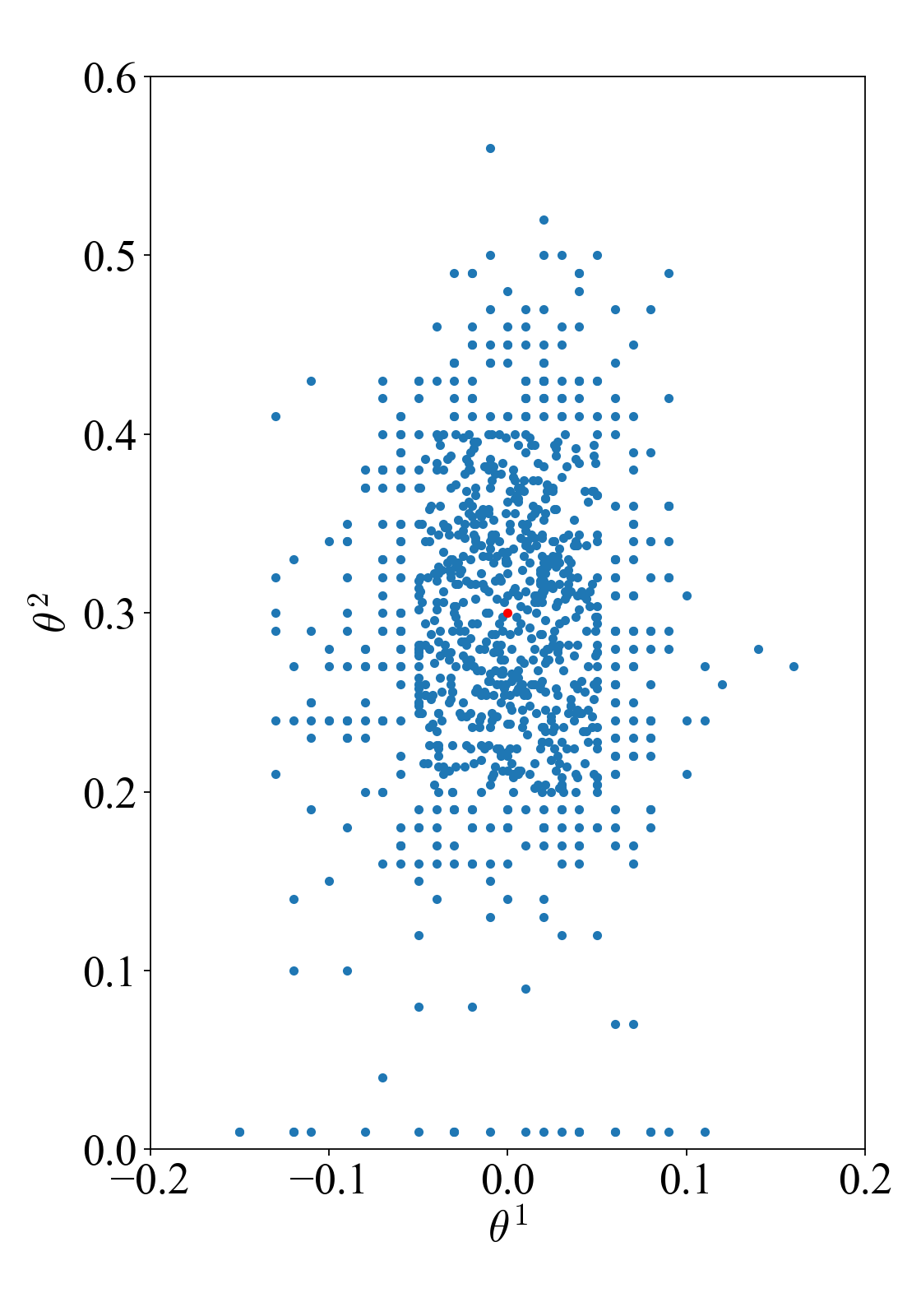}
  \vspace{-6mm}
  \subcaption{$n=1000$}
  \label{fig:mle_0003_1k}
\end{minipage} &
\begin{minipage}{0.24\hsize}
  \centering
  \includegraphics[width=43mm]{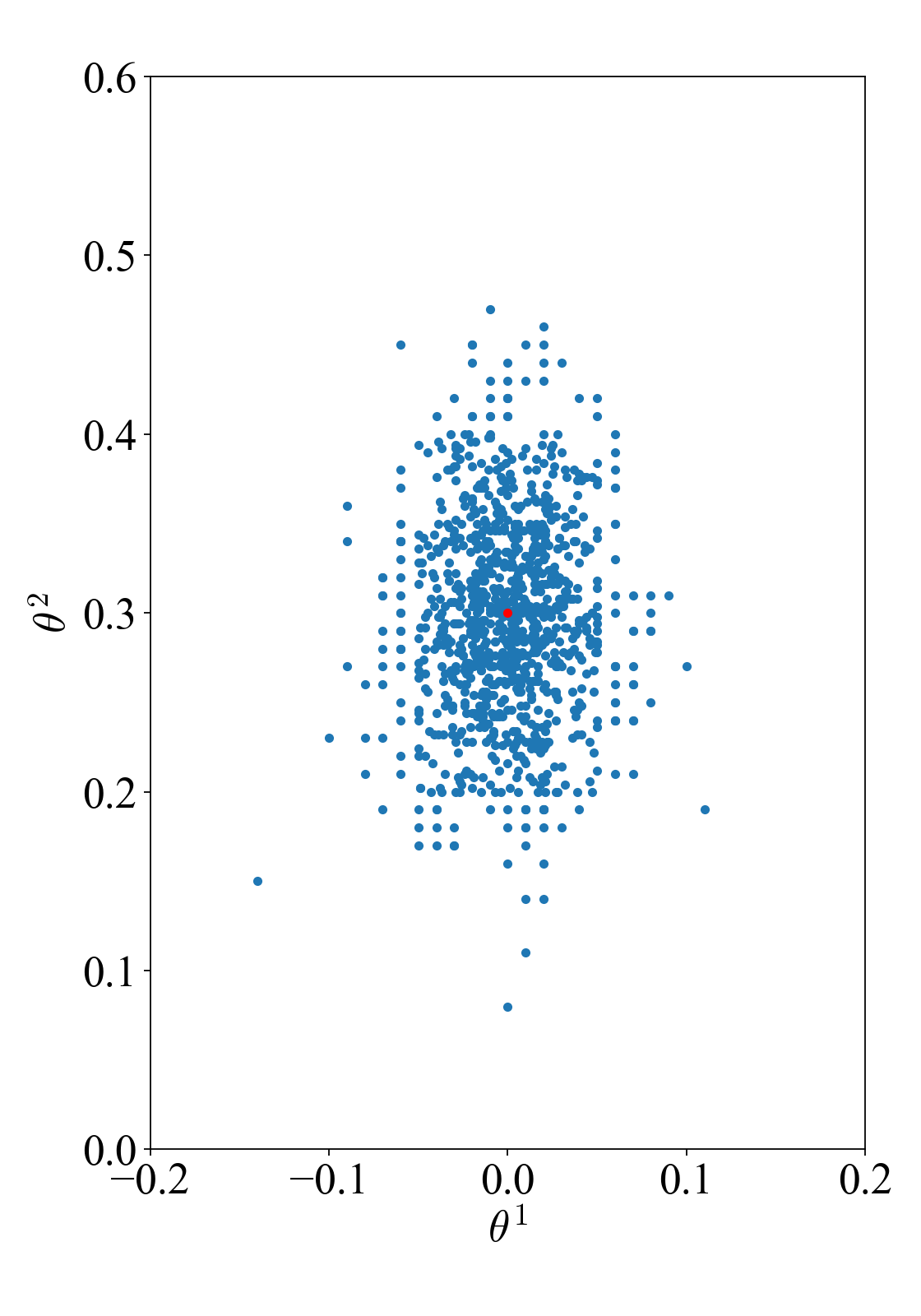}
  \vspace{-6mm}
  \subcaption{$n=2000$}
  \label{fig:mle_0003_2k}
\end{minipage} &
\begin{minipage}{0.24\hsize}
  \centering
  \includegraphics[width=43mm]{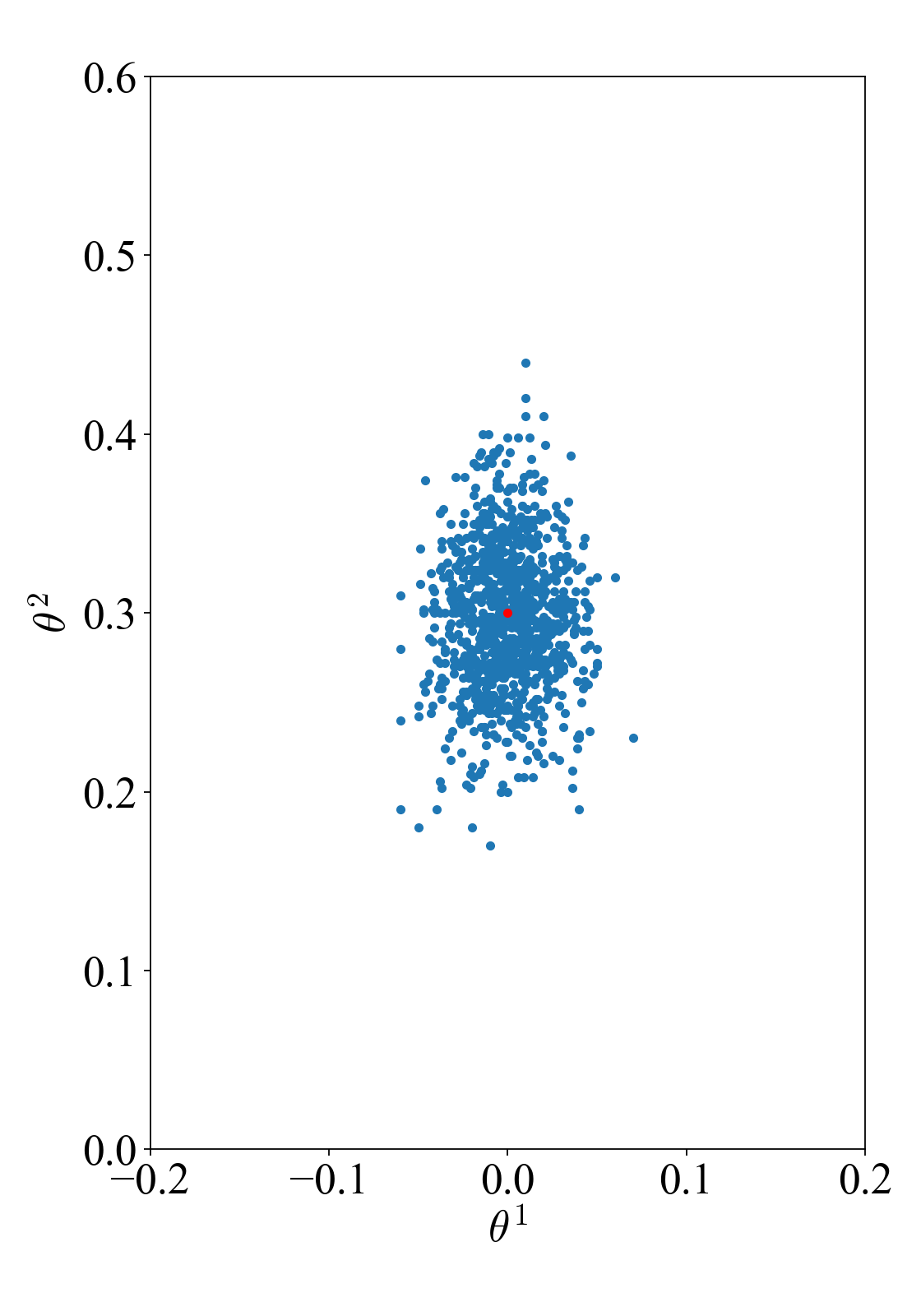}
  \vspace{-6mm}
  \subcaption{$n=4000$}
  \label{fig:mle_0003_4k}
\end{minipage} &
\begin{minipage}{0.24\hsize}
  \centering
  \includegraphics[width=43mm]{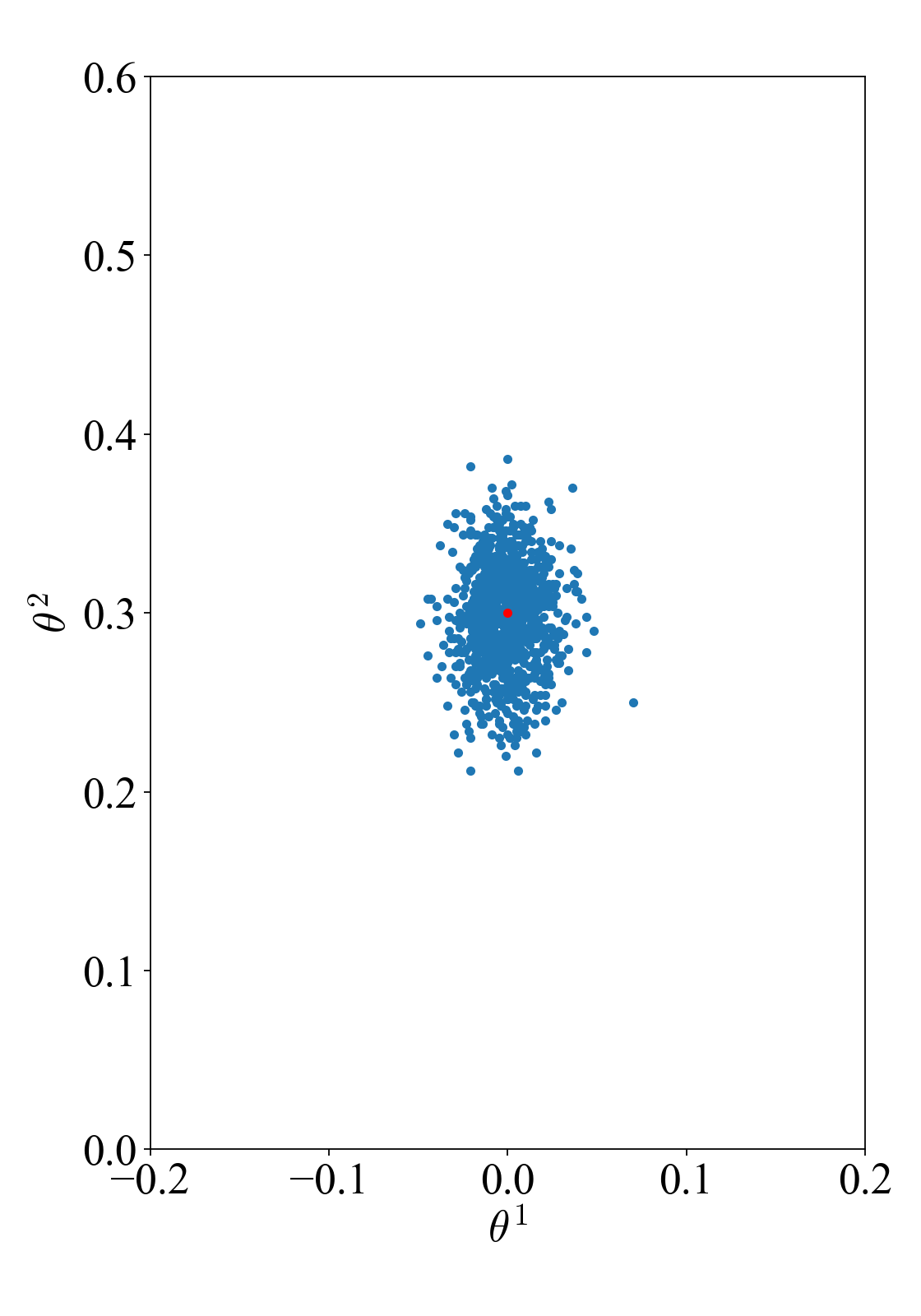}
  \vspace{-6mm}
  \subcaption{$n=8000$}
  \label{fig:mle_0003_8k}
\end{minipage}
\end{tabular}
\captionsetup{singlelinecheck=false, justification=raggedright, font=small}
\caption{Scatter plots of estimated values at several number of steps $n$ when $\theta_*=(\theta_*^1,\theta_*^2)=(0,0.3)$. Each blue dot represents an estimate and the red dot represents the true value of the parameter.}
\label{fig:mle_0003}
\end{figure*}

Next, we confirm that the obtained optimal measurement is superior to the ones used in the previous studies. Fig.~\ref{fig:opt_FI} shows the components of the Fisher information matrix $J_\theta(M({}\cdot{};\theta))$ of the optimal measurement. Since the off-diagonal components are zero, only the diagonal components are shown. The blue curves are the components of the quantum Fisher information matrix $K_\theta$, which gives an upper bound of $J_\theta(M({}\cdot{};\theta))$ as in \eqref{eq:QCR}. It is noteworthy that under the framework of simultaneous estimation of the centroid $\theta^1$ and the separation $\theta^2$, the Fisher information for $\theta^2$ (dashed-and-dotted red curve) converges to a positive value in the limit of $\theta^2\downarrow0$. In other words, it is possible to estimate $\theta^1$ and $\theta^2$ simultaneously with reasonable accuracy no matter how close to zero the separation $\theta^2$ is. This is a remarkable improvement compared with a direct imaging followed by a misaligned SPADE, where the Fisher information for $\theta^2$ falls to zero in the limit of $\theta^2\downarrow0$.

\subsection{Simulating AQSE}

Now we proceed to numerical simulations of AQSE for two point sources using the optimal measurements obtained in the previous subsection. 
In the rest of this paper, we set $\sigma=1$ without loss of generality.

The settings for AQSE are as follows. The true value of the parameter is $\theta_*=(\theta_*^1,\theta_*^2)=(0,0.3)$, the initial estimate is $\hat\theta_0=(1,1)$, and the estimate is computed up to $n=8000$ steps. Since the measurement is updated step by step, it is computationally demanding to obtain a rigorous maximum likelihood estimate. Therefore, the maximum likelihood estimate is approximately obtained by computing the log-likelihood at predefined grid points. Since we want to check the asymptotic behavior of the estimates, the grid points are set finer around the true value of the parameter.

We calculated a sequence of estimates $\hat\theta_1,\hat\theta_2,\hdots,$ $\hat\theta_{8000}\in\Theta$ in each run of AQSE, and repeated such runs 1000 times, to obtain 1000 samples of the sequence of estimates.

First, we check the consistency \eqref{eq:sc}. Fig.~\ref{fig:mle_0003} plots the estimates $\hat\theta_n$ for $n=1000$, 2000, 4000, and 8000. In each figure, the horizontal axis is $\theta^1$ and the vertical axis is $\theta^2$, with the blue dots representing estimates and the red dot representing the true value. The estimates are initially widely scattered around the true value, but as the number of steps increases, the estimates get closer to the true value. 

Next, we check the asymptotic normality \eqref{eq:ae}. We performed goodness-of-fit tests on 1000 samples of $\hat\theta_{8000}$ under the null hypothesis that they follow a multivariate normal distribution. The Anderson--Darling test
in the {\tt mvnTest} package of R 
yielded a p-value of 0.9349, and the Cram\'er--von Mises test in the same package yielded a p-value of 0.9434. 
The null hypothesis was accepted with a very high p-value for both tests.

\begin{figure}[ht]
\begin{tabular}{c}
\begin{minipage}{0.98\hsize}
  \centering
  \includegraphics[width=75mm]{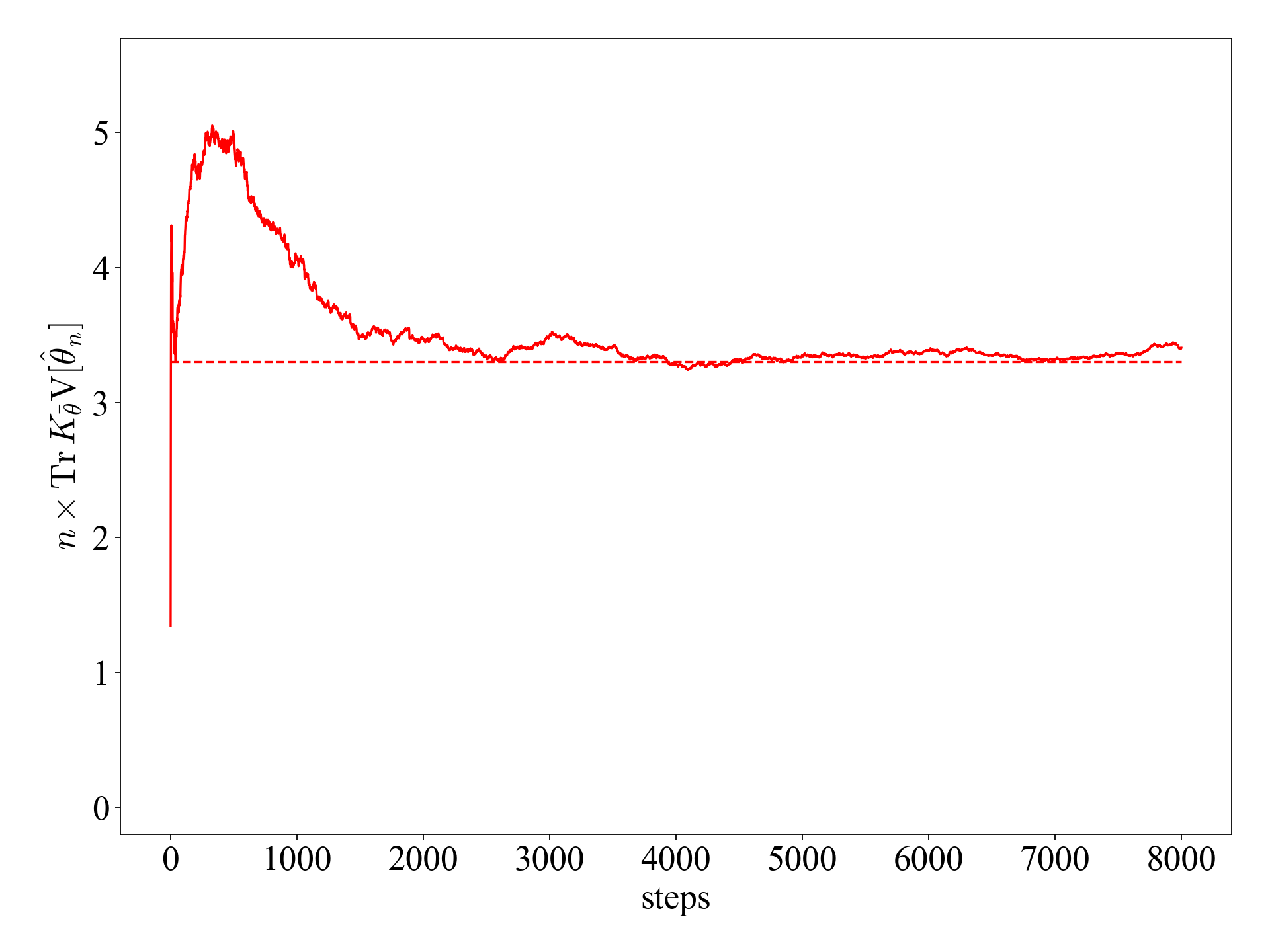}
  \subcaption{$\theta_*=(\theta_*^1,\theta_*^2)=(0,0.3)$}
  \label{fig:scv_0003}
\end{minipage} \\[32mm]
\begin{minipage}{0.98\hsize}
  \centering
  \includegraphics[width=75mm]{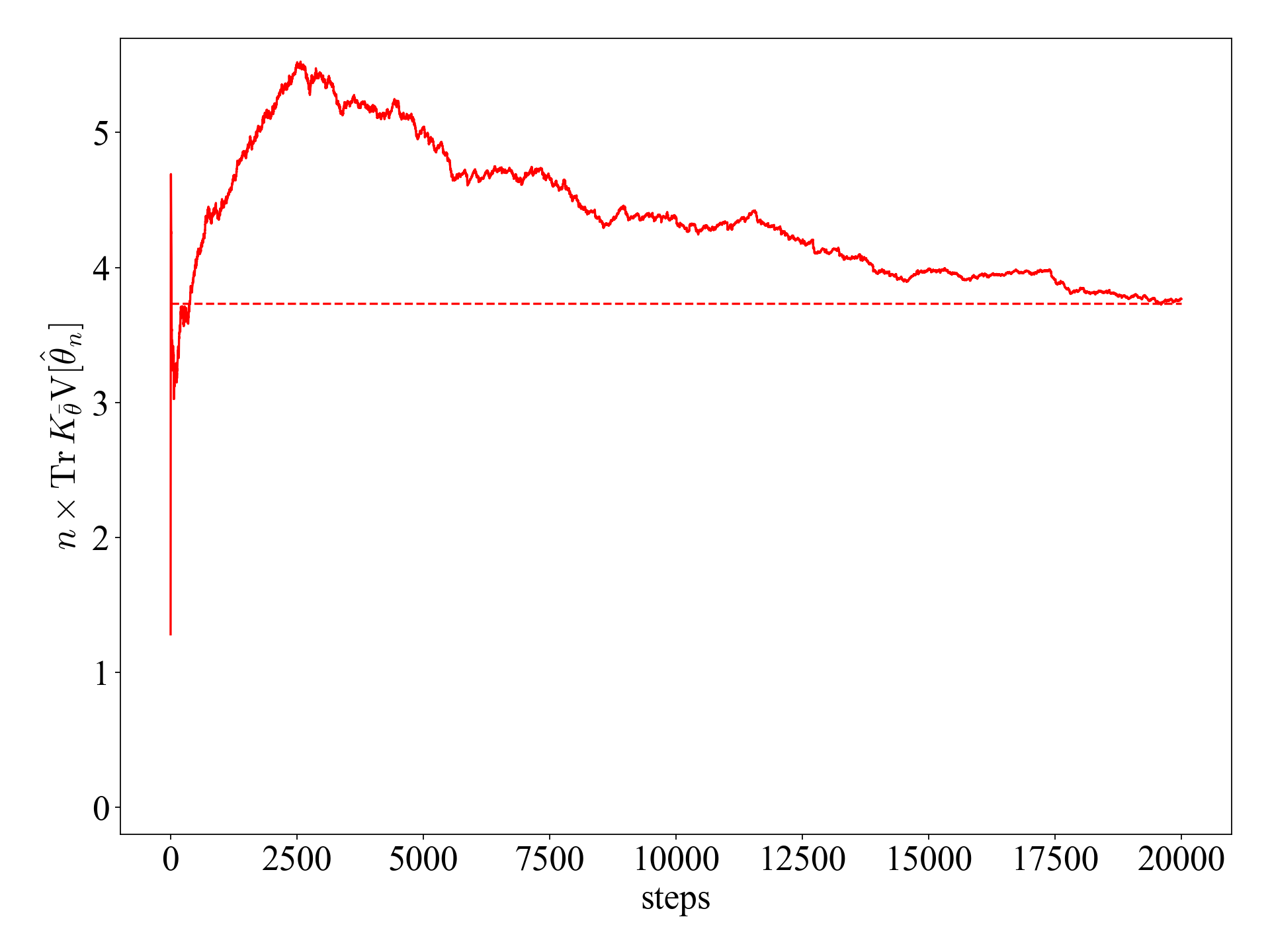}
  \subcaption{$\theta_*=(\theta_*^1,\theta_*^2)=(0,0.1)$}
  \label{fig:scv_0001}
\end{minipage}
\end{tabular}
\captionsetup{singlelinecheck=false, justification=raggedright, font=small}
\caption{Weighted trace of the sample covariance matrix for (a) $\theta_*=(\theta_*^1, \theta_*^2)=(0, 0.3)$ and (b) $\theta_*=(\theta_*^1, \theta_*^2)=(0, 0.1)$. The dashed lines indicate theoretical limits given in Fig.~\ref{fig:opt_v}. 
Comparing (a) and (b), we find a notable reduction in the rate of convergence of sample covariance as $\theta_*^2$ approaches zero.}
\label{fig:scv}
\end{figure}

Finally, we check how the sample covariance matrix $V[\hat\theta_n]$ evolves with the number of steps. Fig.~\ref{fig:scv_0003} shows the weighted trace of the sample covariance matrix $\tr K_{\bar\theta}V[\hat\theta_n]$, where $K_{\bar\theta}$ is the quantum Fisher information matrix at the sample mean $\bar\theta$ of the estimates at each step. For the sake of comparison, Fig.~\ref{fig:scv_0001} also shows the result for the case where the true value is $\theta_*=(\theta_*^1,\theta_*^2)=(0,0.1)$. Note that the values of the weighted trace are multiplied by $n$, since the sample covariance matrix decreases by $1/n$. The dashed lines indicate the ultimate limits of estimation precision displayed in Fig.~\ref{fig:opt_v}. In each case, the solid curve approaches the dashed line as the number of steps increases. This means that if the number of steps is large enough, we can estimate the centroid $\theta^1$ and the separation $\theta^2$ simultaneously with the best accuracy theoretically possible.

\subsection{Trapping phenomena near \texorpdfstring{$\theta^2=0$}{}}

\iftrue
\begin{figure*}[tb]
\begin{tabular}{c}

\begin{minipage}{0.99\hsize}
  \centering
  \includegraphics[width=45mm]{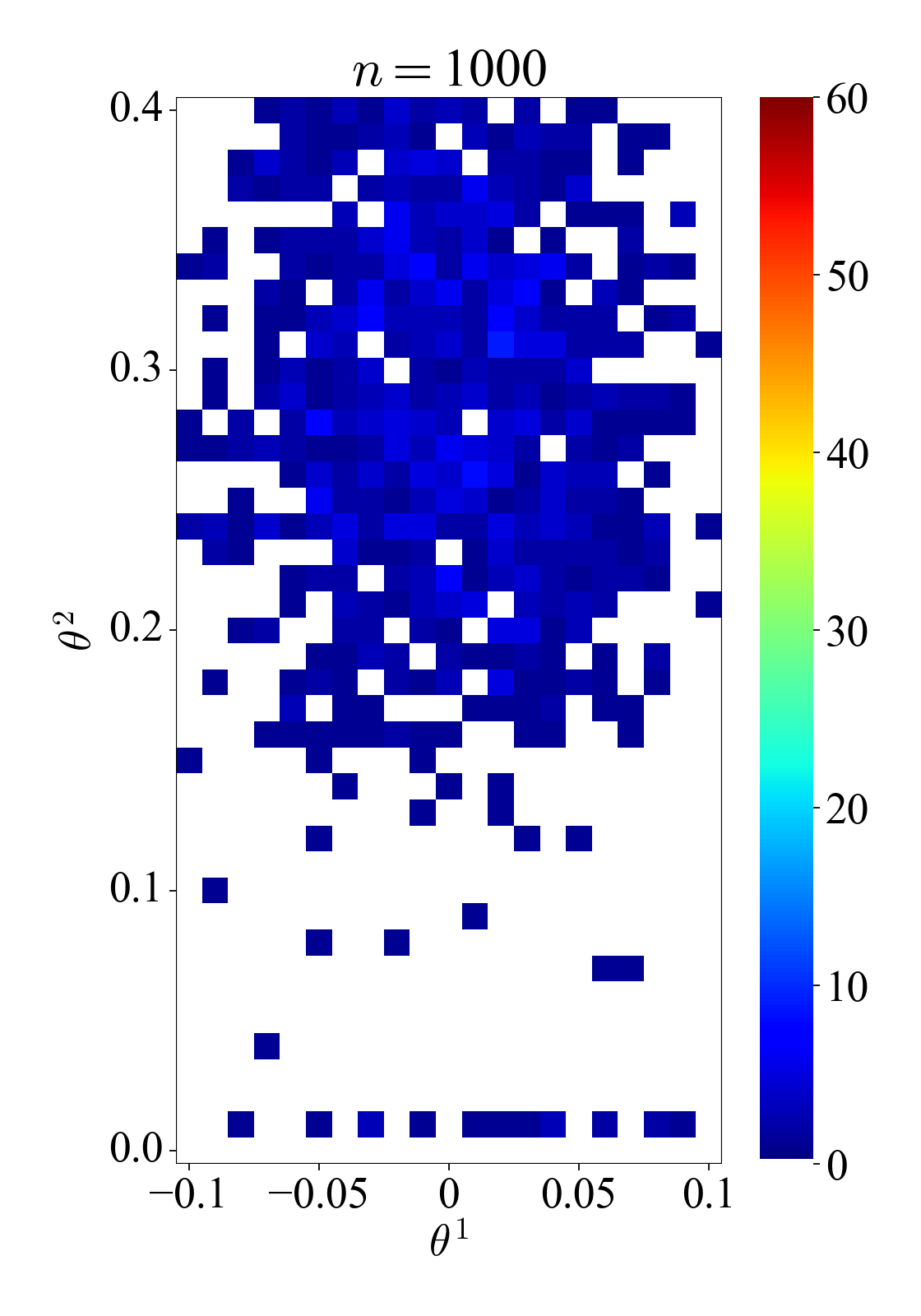}
  \hspace{-4mm}
  \includegraphics[width=45mm]{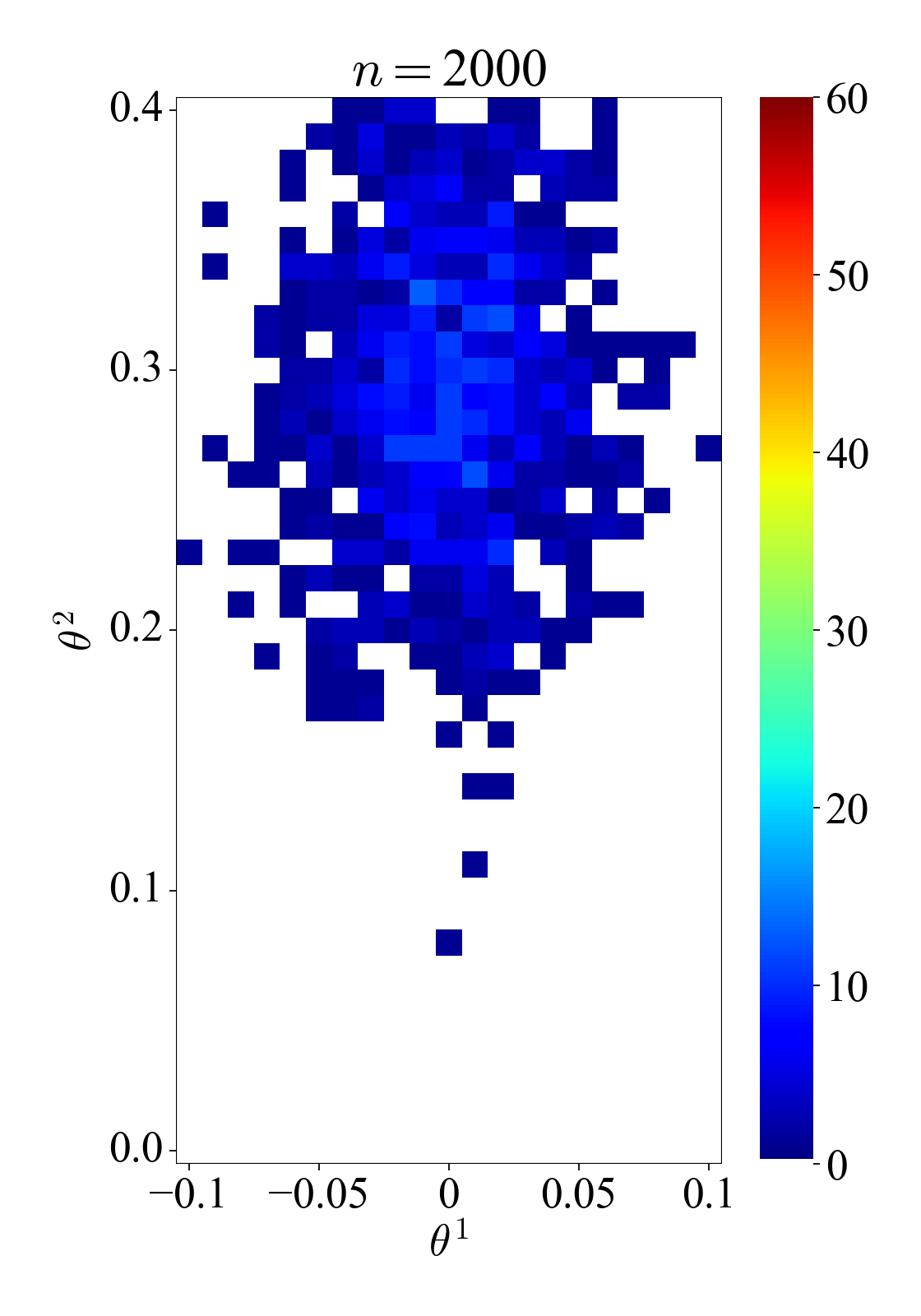}
  \hspace{-4mm}
  \includegraphics[width=45mm]{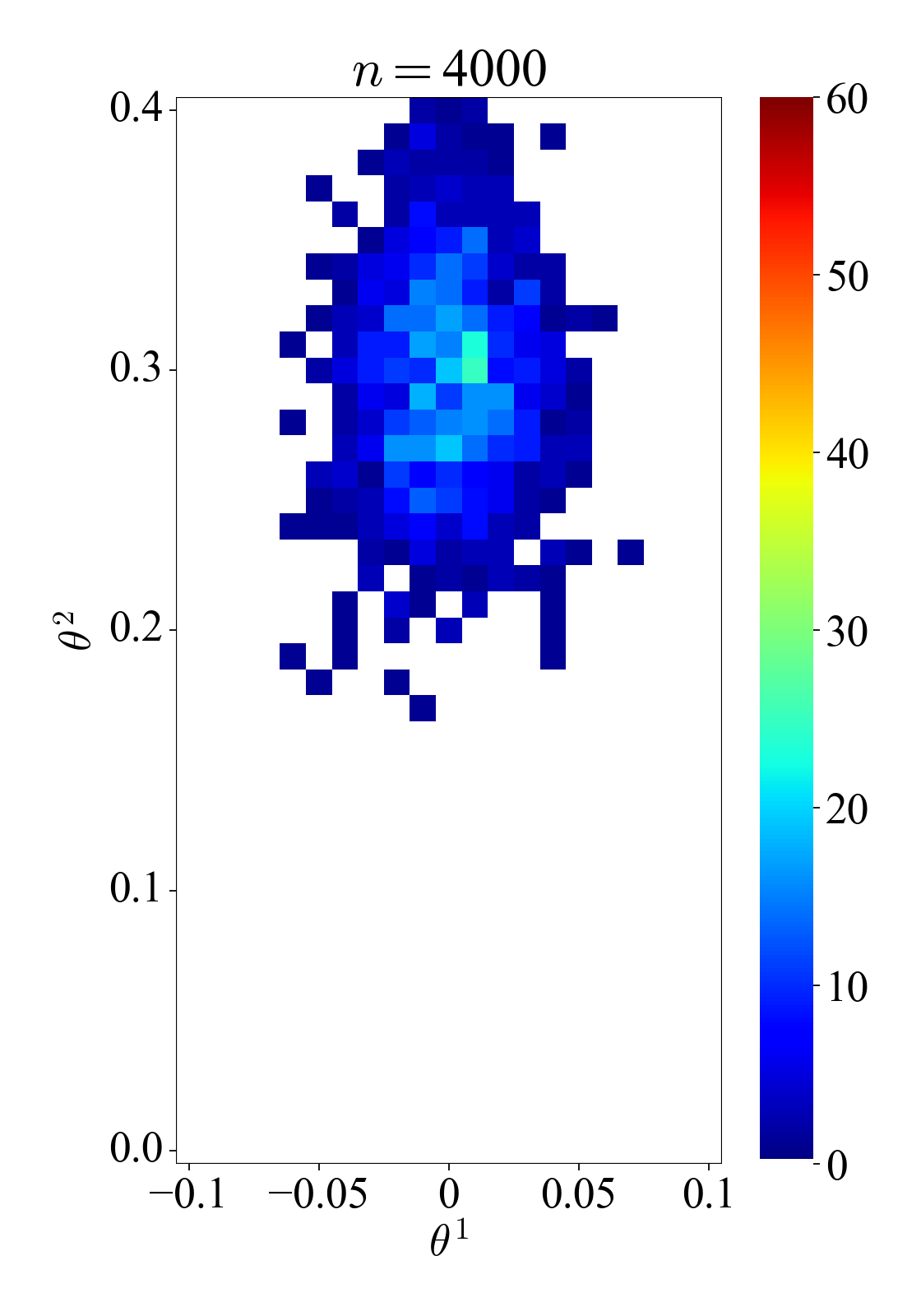}
  \hspace{-4mm}
  \includegraphics[width=45mm]{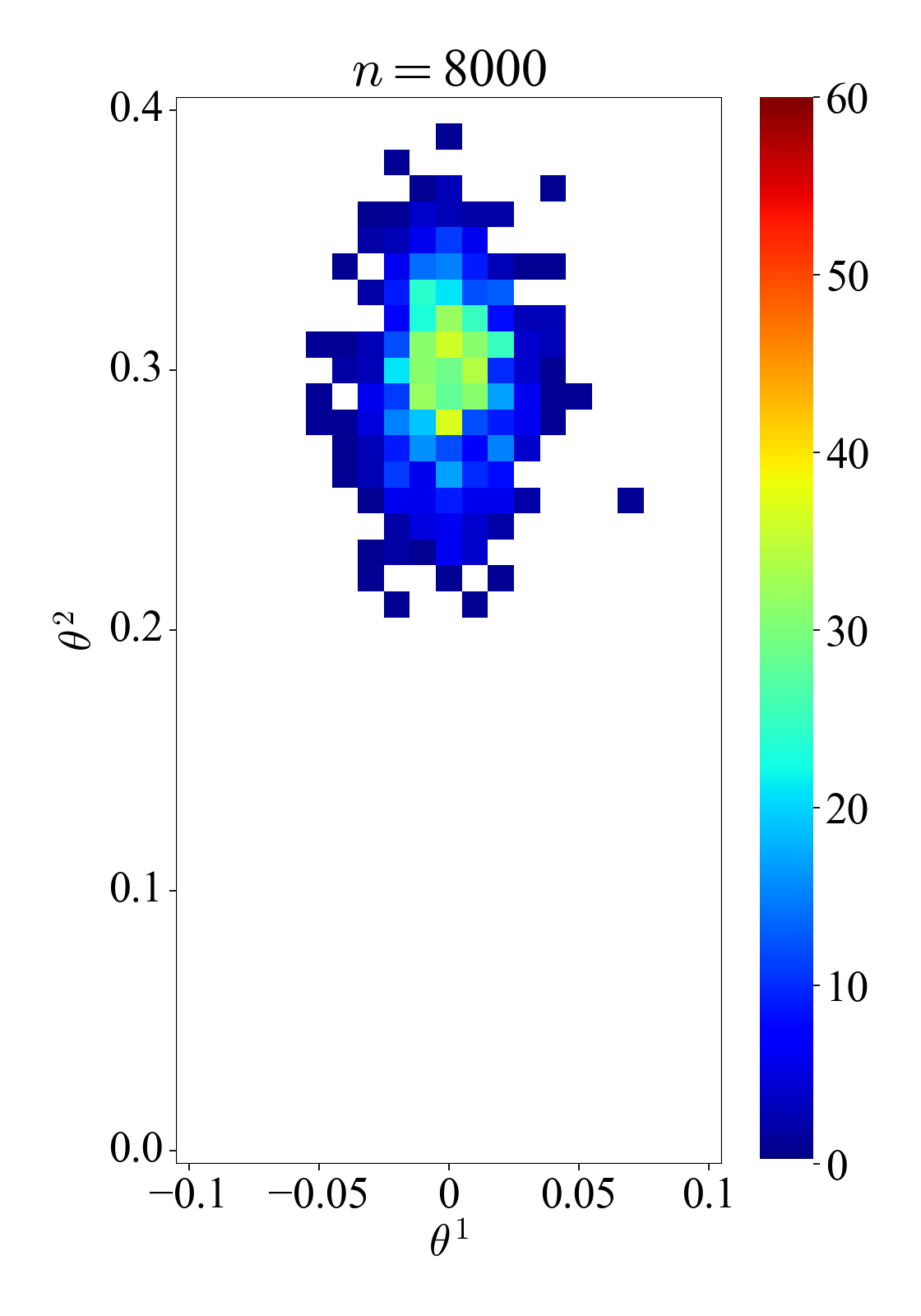}
  \vspace{-3mm}
  \subcaption{$\theta_*=(\theta_*^1,\theta_*^2)=(0,0.3)$}
\end{minipage} \\

\begin{minipage}{0.99\hsize}
  \centering
  \includegraphics[width=45mm]{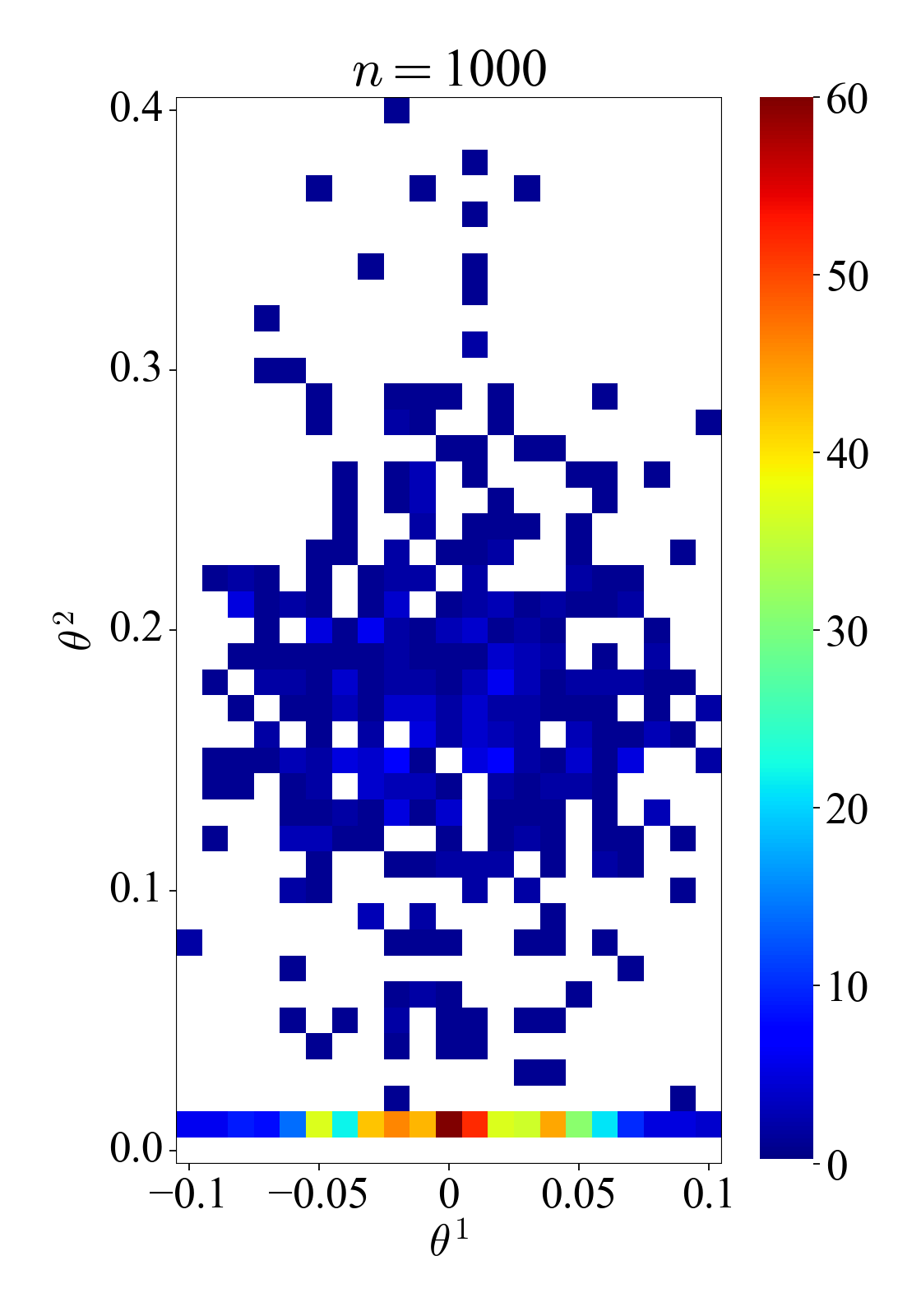}
  \hspace{-4mm}
  \includegraphics[width=45mm]{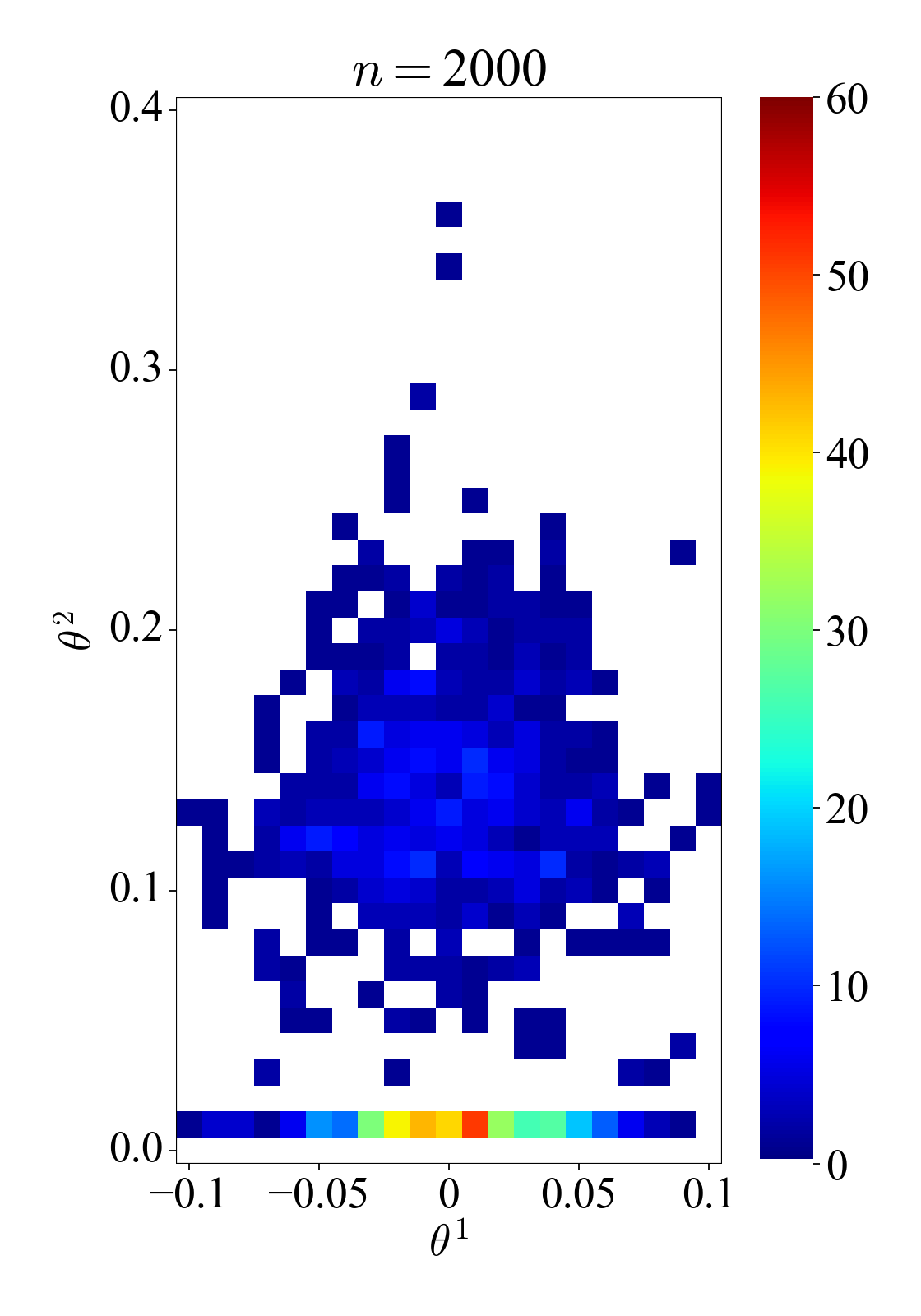}
  \hspace{-4mm}
  \includegraphics[width=45mm]{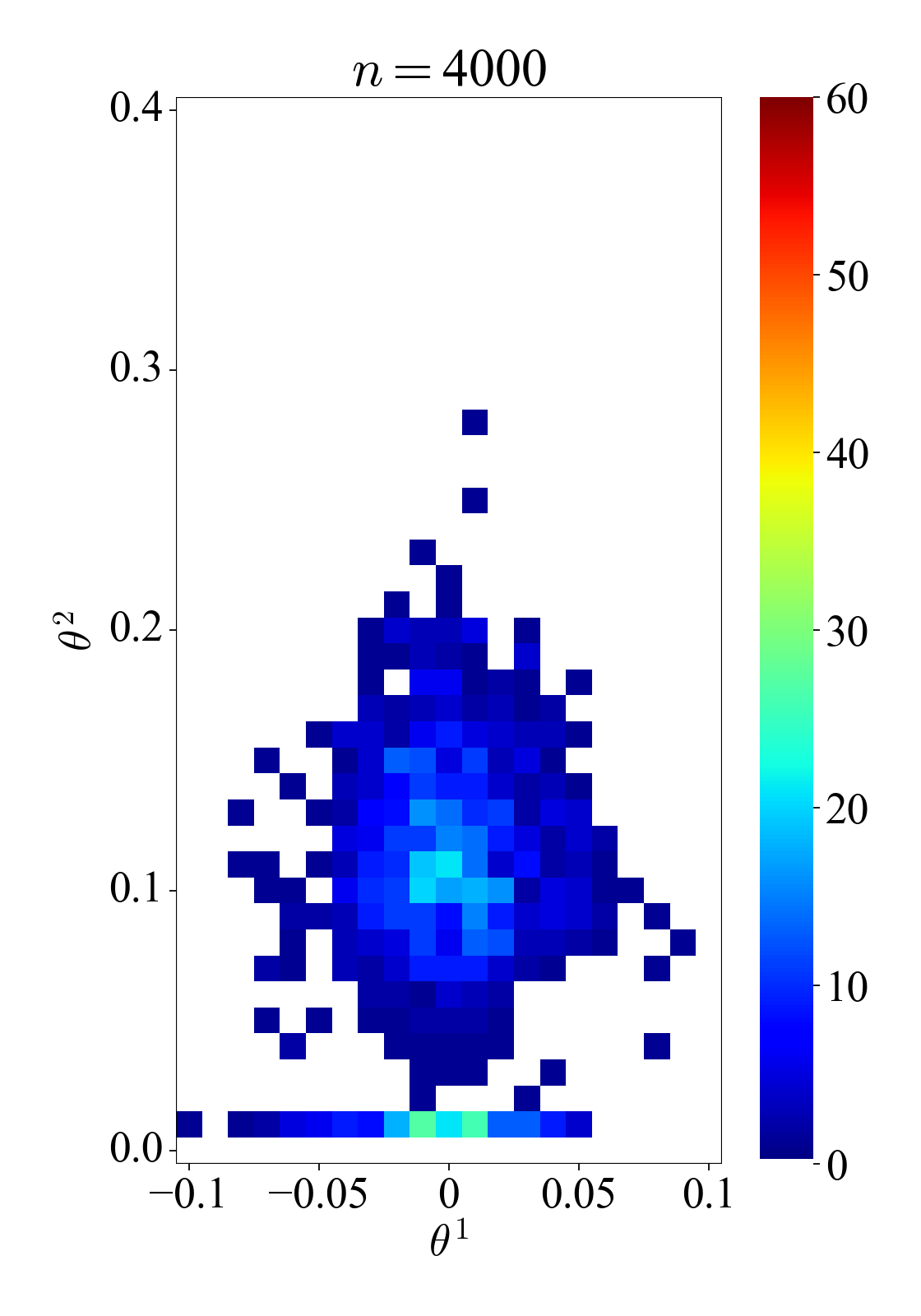}
  \hspace{-4mm}
  \includegraphics[width=45mm]{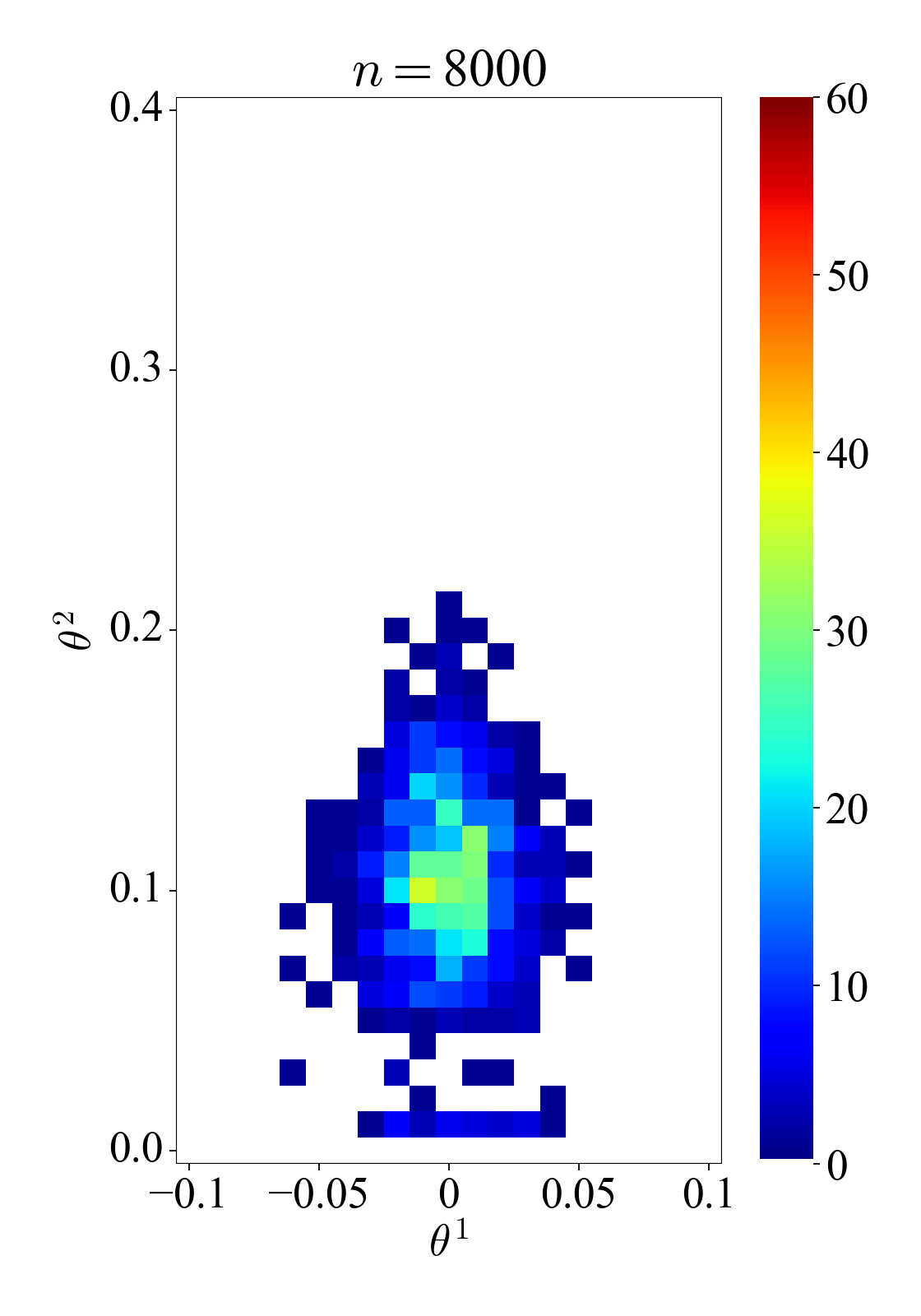}
  \vspace{-3mm}
  \subcaption{$\theta_*=(\theta_*^1,\theta_*^2)=(0,0.1)$}
\end{minipage}

\end{tabular}
\captionsetup{singlelinecheck=false, justification=raggedright, font=small}
\caption{Heatmaps of estimates at several number of steps $n$ for (a) $\theta_*=(\theta_*^1,\theta_*^2)=(0,0.3)$ and (b) $\theta_*=(\theta_*^1,\theta_*^2)=(0,0.1)$. 
Comparing the upper and lower panels, we see that when $\theta_*^2$ is small, a significant number of estimates are trapped near the boundary $\theta^2=0$ for a long time.}
\label{fig:heatmaps}
\end{figure*}

\else

\begin{figure*}[tb]
\begin{tabular}{cccc}
\begin{minipage}{0.24\hsize}
  \centering
  \includegraphics[width=43mm]{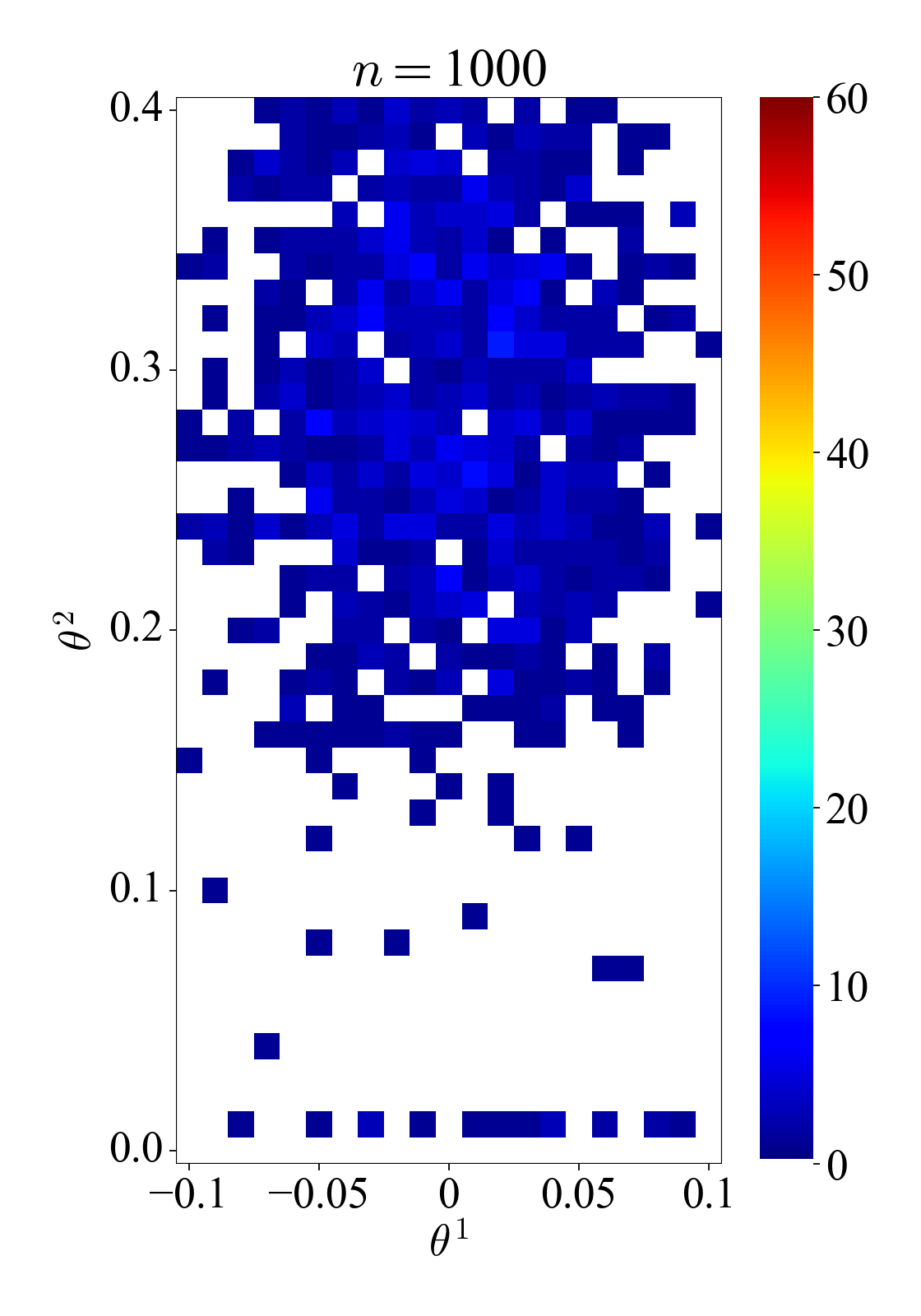}
  \vspace{-6mm}
  \subcaption{$n=1000$}
\end{minipage} &
\begin{minipage}{0.24\hsize}
  \centering
  \includegraphics[width=43mm]{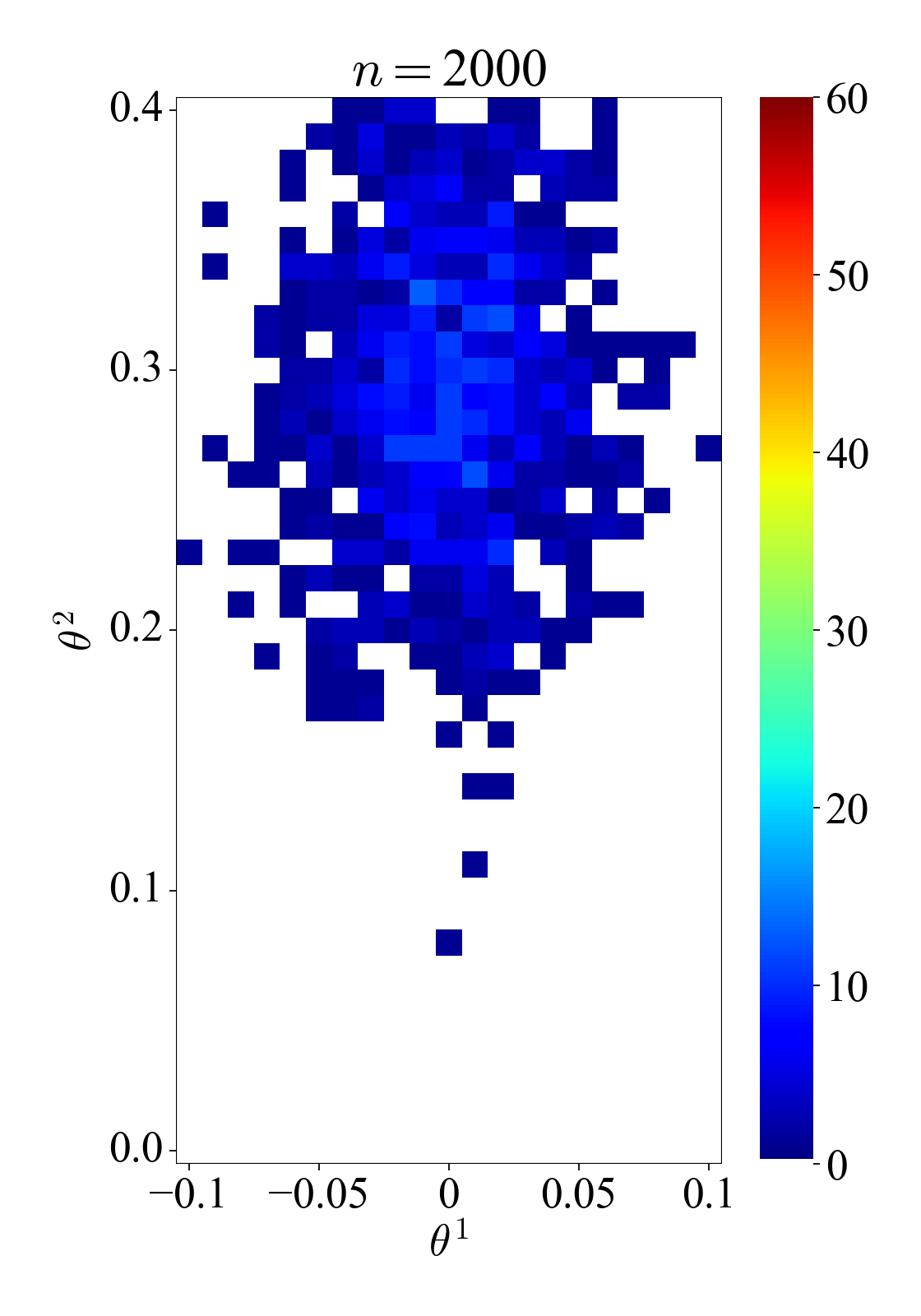}
  \vspace{-6mm}
  \subcaption{$n=2000$}
\end{minipage} &
\begin{minipage}{0.24\hsize}
  \centering
  \includegraphics[width=43mm]{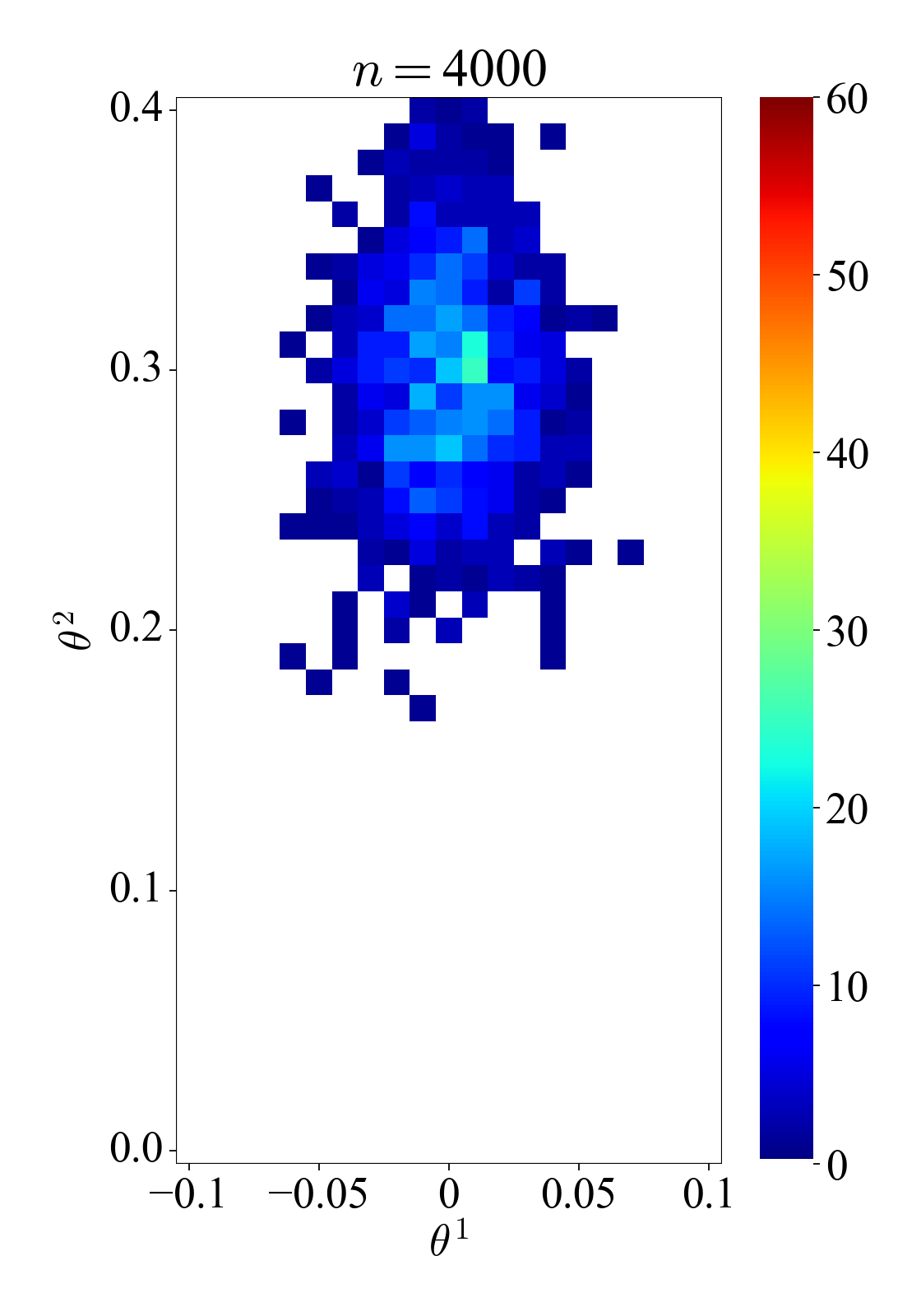}
  \vspace{-6mm}
  \subcaption{$n=4000$}
\end{minipage} &
\begin{minipage}{0.24\hsize}
  \centering
  \includegraphics[width=43mm]{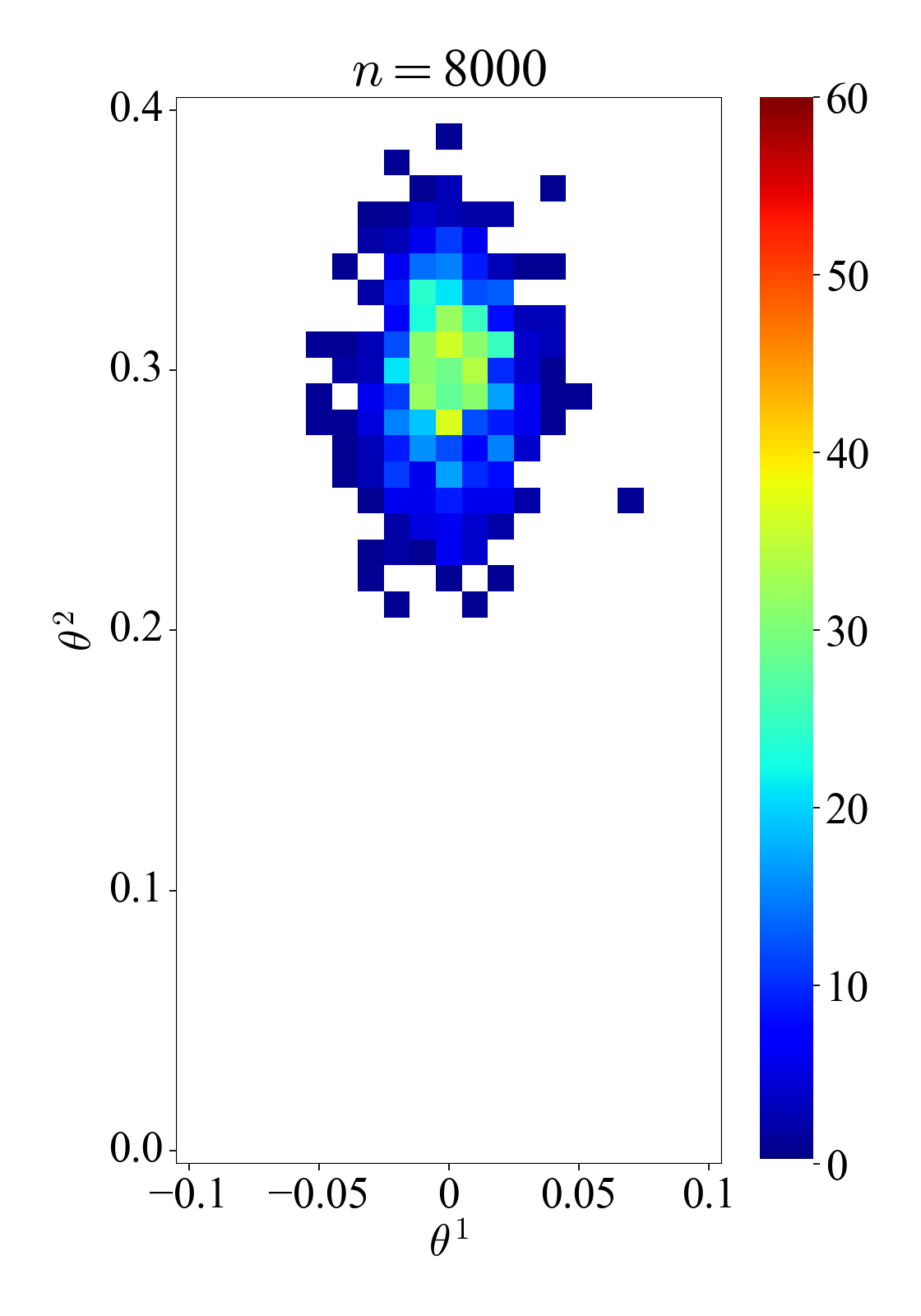}
  \vspace{-6mm}
  \subcaption{$n=8000$}
\end{minipage} \\[31mm]

\begin{minipage}{0.24\hsize}
  \centering
  \includegraphics[width=43mm]{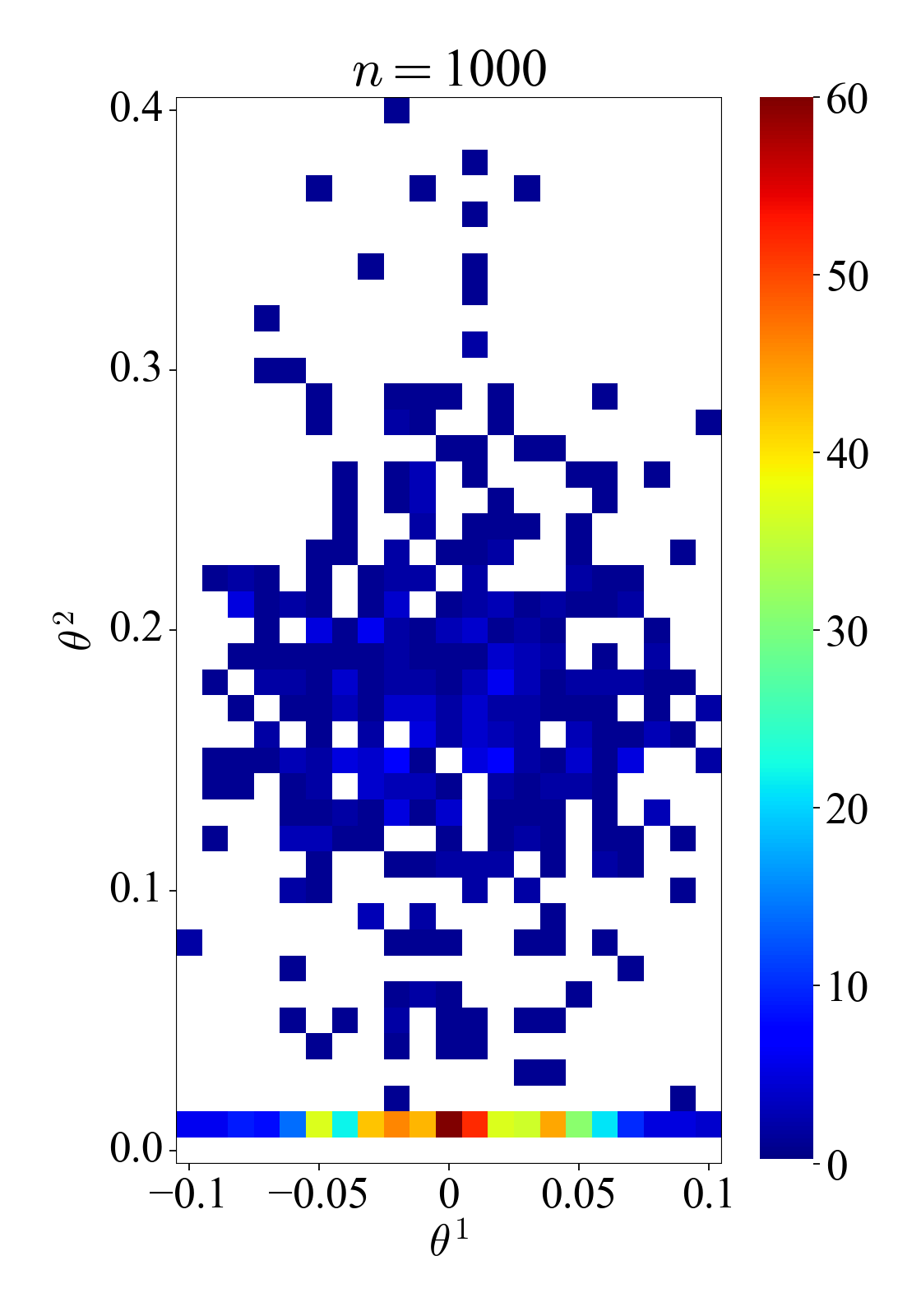}
  \vspace{-6mm}
  \subcaption{$n=1000$}
\end{minipage} &
\begin{minipage}{0.24\hsize}
  \centering
  \includegraphics[width=43mm]{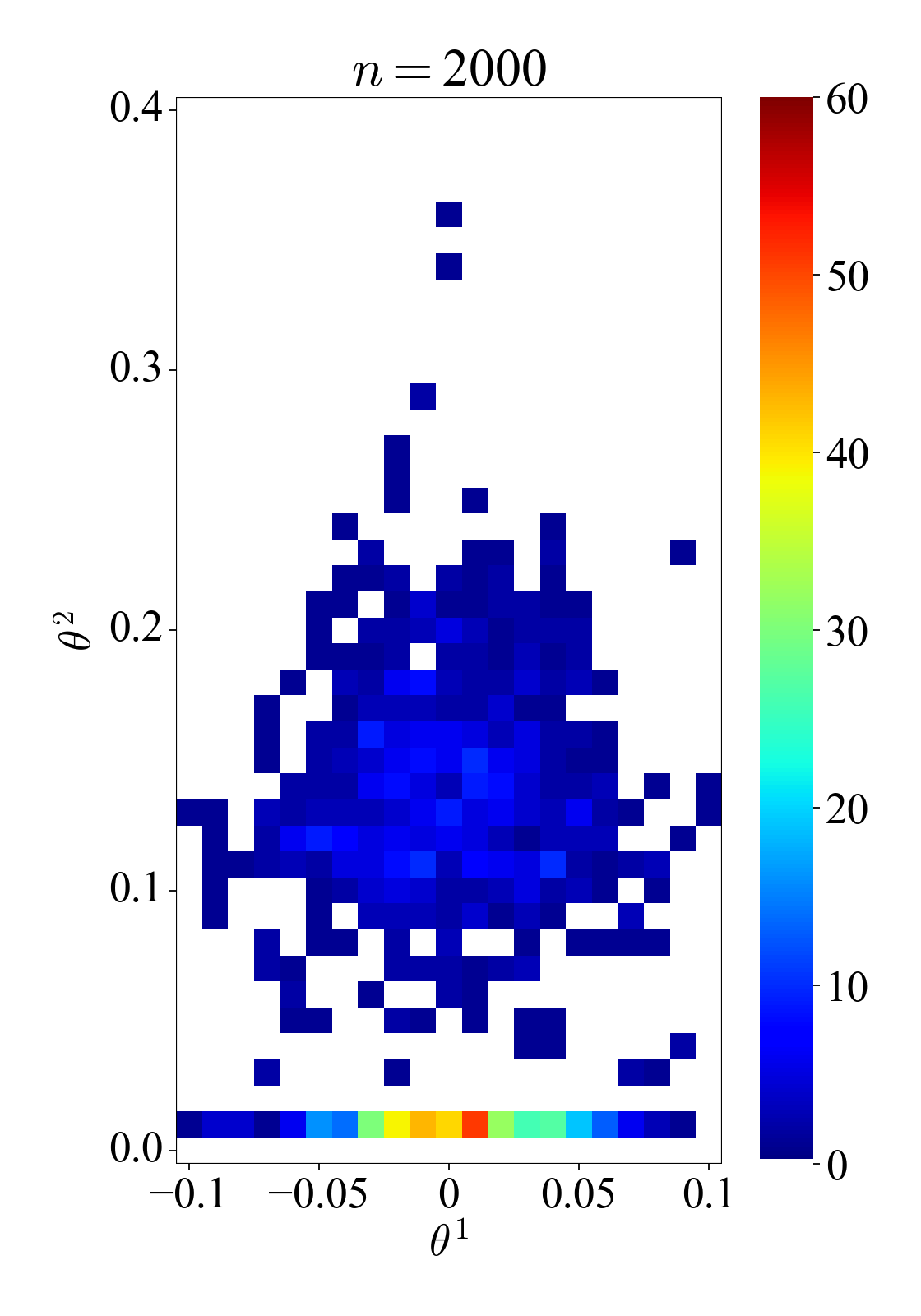}
  \vspace{-6mm}
  \subcaption{$n=2000$}
\end{minipage} &
\begin{minipage}{0.24\hsize}
  \centering
  \includegraphics[width=43mm]{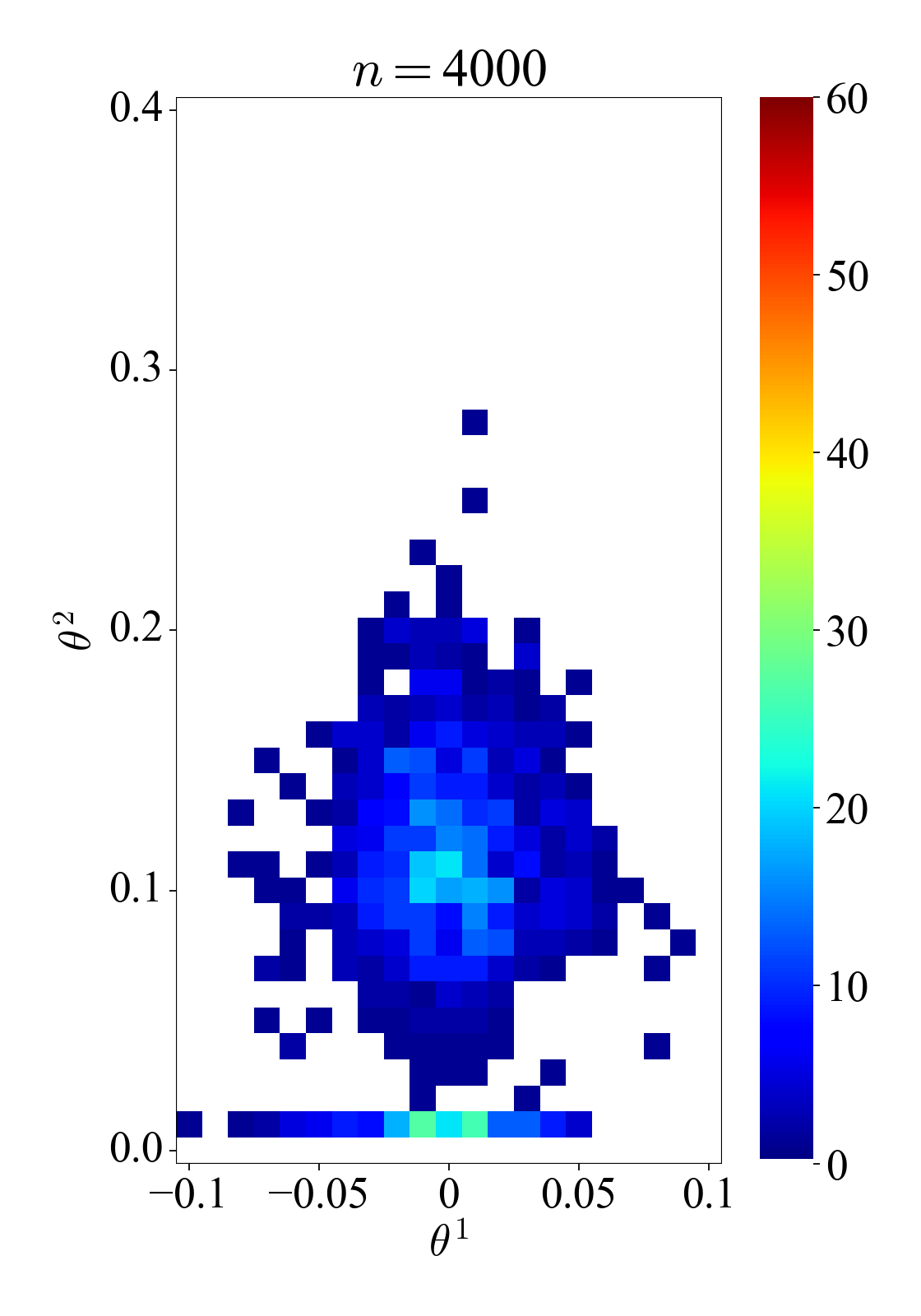}
  \vspace{-6mm}
  \subcaption{$n=4000$}
\end{minipage} &
\begin{minipage}{0.24\hsize}
  \centering
  \includegraphics[width=43mm]{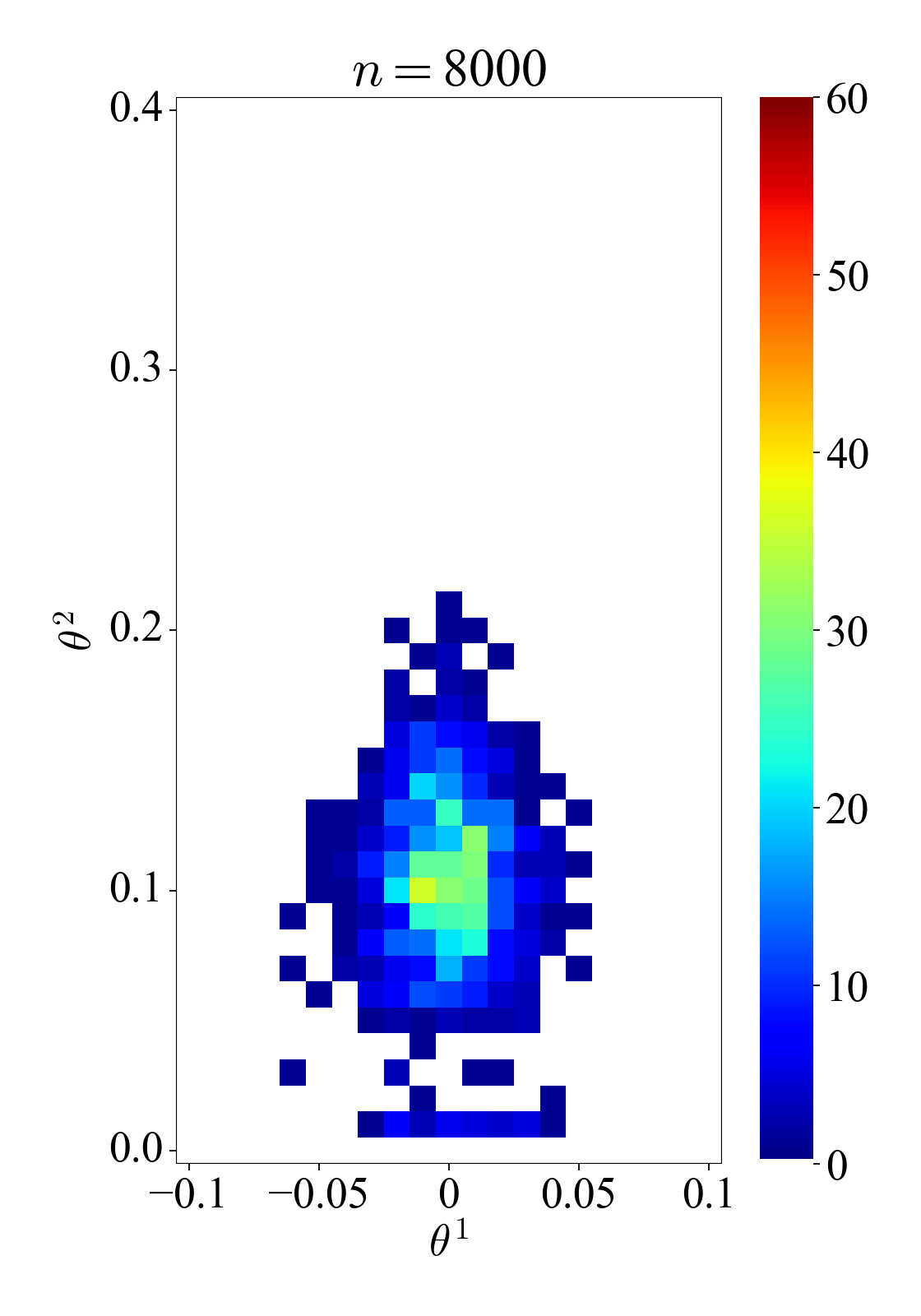}
  \vspace{-6mm}
  \subcaption{$n=8000$}
\end{minipage}
\end{tabular}
\captionsetup{singlelinecheck=false, justification=raggedright, font=small}
\caption{Heatmaps of MLEs at several steps $n$ for (a)--(d) $\theta_*=(\theta_*^1,\theta_*^2)=(0,0.3)$ and (e)--(h) $\theta_*=(\theta_*^1,\theta_*^2)=(0,0.1)$.}
\label{fig:heatmaps}
\end{figure*}
\fi

It is noteworthy that, as can be seen from Figs.~\ref{fig:scv_0003} and \ref{fig:scv_0001}, the convergence becomes much slower as $\theta_*^2$ gets closer to zero.
We also find from the heatmaps in Fig.~\ref{fig:heatmaps}
that, when $\theta_*^2$ is small, a significant number of MLEs are trapped near the boundary $\theta^2=0$ for a long time. 
These observations prompt us to envisage the following scenario: 
when $\theta_*^2$ is small, a good number of estimates are located in the boundary region $\theta^2\approx0$ at an early stage of AQSE because of the large sample dispersion, and are kept trapped in that region for a long time, yielding a notable slowdown of the convergence of the sample covariance matrix.

Let us examine the validity of this ``boundary effect'' scenario by means of the following tentative evaluation: 
because of the nature of convergence in distribution, each contour of the probability density function would converge to $\theta_*$ in the rate $\sim 1/\sqrt{n}$, so that the time $\tau$ for a certain contour to pass through the ``trapping wall'', i.e. the grid line closest to the axis $\theta^2=0$, at a distance $d$ from the true parameter $\theta_*$ may be evaluated as
\begin{equation}
 d\sim \frac{C}{\sqrt{\tau}}\quad\Longleftrightarrow\quad \tau\sim\frac{C^2}{d^2},
\end{equation}
where $C$ is a certain constant corresponding to the contour that characterizes the trapping effect.
Assume further that, after getting out of the influence of the trapping wall, the time $t_0$ required for the estimates to converge in distribution is independent of $\theta_*$.
Then the total time $T=\tau+t_0$ of convergence in distribution would be roughly evaluated as
\begin{equation} \label{eqn:scaling}
 T\sim \frac{C^2}{(\theta_*^2)^2}+t_0.
\end{equation}

\renewcommand\ttdefault{SourceCodePro-TLF}
\renewcommand\bfdefault{sb}

\begin{figure}[ht]
  \centering
  \includegraphics[width=85mm]{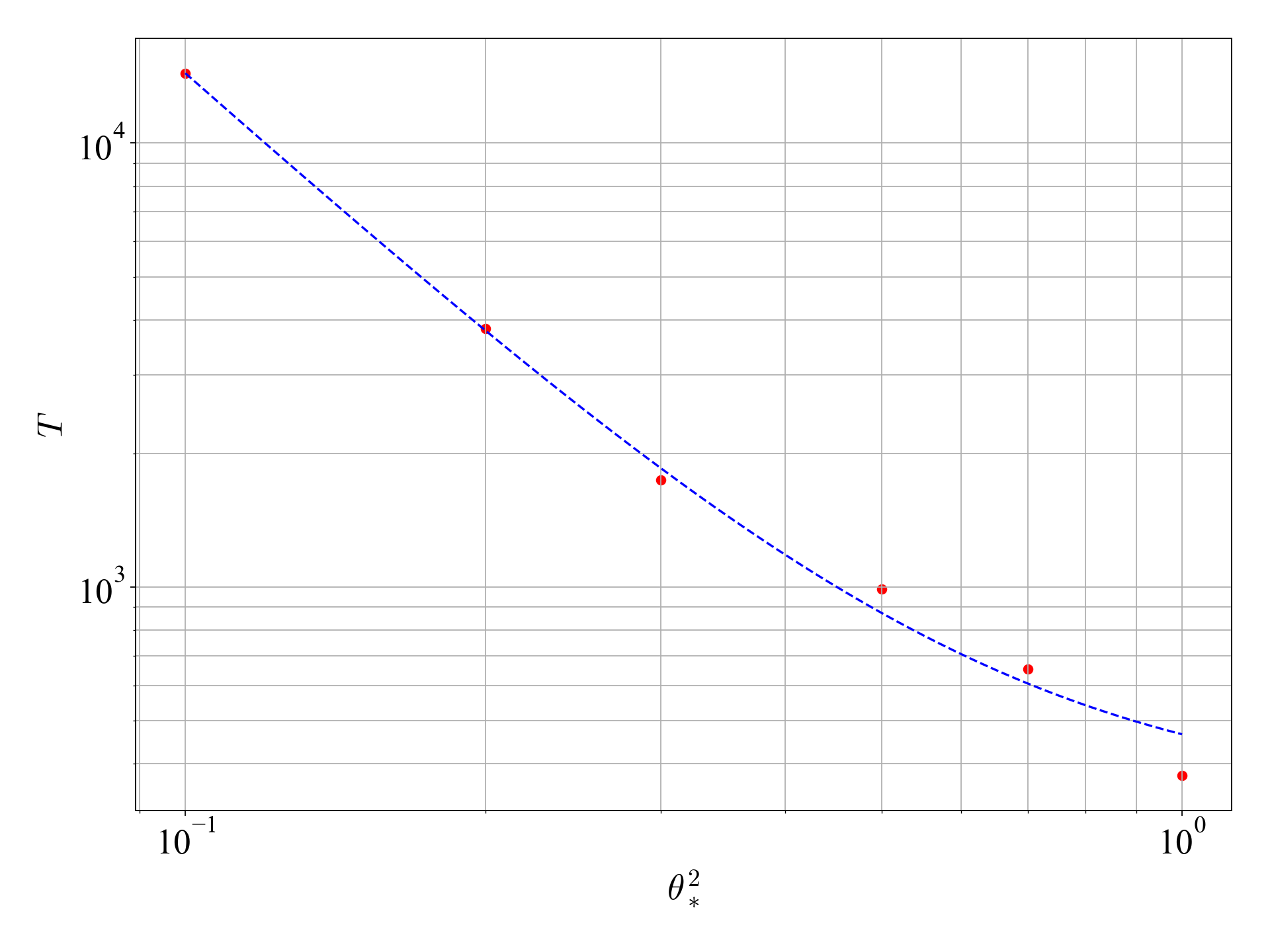}
\captionsetup{singlelinecheck=false, justification=raggedright, font=small}
\caption{Time $T$ required for the estimates to converge in distribution for several values of $\theta_*^2$. 
The data are fitted by the curve $T={c}/{(\theta_*^2)^e}+t_0$ with $e=2.02$, $c=133$, and $t_0=333$.}
\label{fig:scaling}
\end{figure}

Let us verify the validity of this scaling law. 
The following is a list of convergence time $T$ obtained by numerical simulations for several values of $\theta_*^2$: 
\begin{alignat*}{3}
 (\theta_*^2, T)
 =\;&(0.1, 14372),&\; & (0.2, 3816),&\; & (0.3, 1744), \\
  &(0.5, 990),& & (0.7, 653),& & (1.0, 376);
\end{alignat*}
the first and the third data corresponding to Figs.~\ref{fig:scv_0001} and \ref{fig:scv_0003}, respectively. 
Here, we take $T$ as the first time at which the weighted trace of the sample covariance matrix decreases to within 5\% of the theoretical limit.
A nonlinear parameter fitting using {\tt\bfseries FindFit} function of {\it Mathematica} yields
\begin{equation}
 T=\frac{c}{(\theta_*^2)^e}+t_0
\end{equation}
with $e=2.02$, $c=133$, and $t_0=333$ as seen in Fig.~\ref{fig:scaling}.
This result is reasonably consistent with the scaling law \eqref{eqn:scaling}, supporting the validity of the trapping scenario.

\renewcommand\ttdefault{cmtt}
\renewcommand\bfdefault{bx}

In summary, although the centroid $\theta_*^1$ and the separation $\theta_*^2$ can in principle be estimated simultaneously with the best accuracy in the asymptotic limit, a notable reduction in the rate of convergence of estimates arises as the separation $\theta_*^2$ gets closer to zero.
The slowdown of the convergence of estimates as $\theta_*^2 \downarrow 0$ may be regarded as a manifestation of Rayleigh's curse in the quantum domain.

\section{Conclusion} \label{sec:summary}
In this paper, we proposed a method to estimate the centroid $\theta^1$ and the separation $\theta^2$ of two point sources simultaneously by AQSE. 
Numerical experiments have confirmed that the method works properly if the number of steps is large enough. 
It was also found that the closer to zero the $\theta^2$ component of the true value is, the slower the convergence of estimates becomes. 
This phenomenon may suggest that Rayleigh's curse may still survive in the framework of quantum theory, transforming itself into a {\it plateau} phenomenon, a notable reduction in the rate of convergence  of estimates in AQSE.
Nevertheless, the mechanism behind the plateau phenomenon requires further investigation.

\appendix*

\section{Numerical optimization of measurement} \label{append}

In this Appendix, we explain how to reduce the problem of finding the optimal measurement $M({}\cdot{};\theta)$ given by \eqref{eq:opt_m} to an unconstrained nonlinear programming problem \cite{yamagata_m}. 

As described in Sec.~\ref{subsec:opt_meas}, $\rho_\theta,\,\f{\pd\rho_\theta}{\pd\theta^1}$ and $\f{\pd\rho_\theta}{\pd\theta^2}$ appearing in the objective function have support in the $\theta$-dependent four-dimensional subspace $\mcal{V}_\theta$, so $M({}\cdot{};\theta)$ can be obtained as a POVM on $\mcal{V}_\theta$. Since the model is a real model, only the real part of the POVM need to be considered. In addition, since the Fisher information matrix does not become smaller by decomposing the POVM into rank-one measurement, it is sufficient to consider only rank-one measurements. Furthermore, according to Fujiwara \cite{fujiwara06}, the optimal measurement can be achieved with at most 16-valued measurement. Originally, he stated that $(\dim\mcal{V}_\theta)^2+d(d+1)$ is sufficient for the number of measurement outcomes where $d$ is the dimension of the parameter $\theta$, but $(\dim\mcal{V}_\theta)^2$ can be replaced by $\f12\dim\mcal{V}_\theta(\dim\mcal{V}_\theta+1)$ since our model is a real model.
 
We now consider the parametrization of an $n$-valued real rank-one measurement on a $q$-dimensional Hilbert space $\mathbb{C}^q$. The $n$-valued real rank-one measurement is given by real vectors $a_1,a_2,\hdots,a_n\in\mathbb{C}^q$ satisfying
\begin{equation} \label{eq:rank1m}
\sum_{i=1}^n \ket{a_i}\bra{a_i} = I_q.
\end{equation}
Although $a_1,a_2,\hdots,a_n$ must satisfy the above constraint, real rank-one measurements can be parametrized without any constraint as follows \cite{yamagata_m}.

\eqref{eq:rank1m} can be rewritten as
\begin{align}
\lt(\begin{array}{cccc} a_1 & a_2 & \cdots & a_n \end{array}\rt)
\lt(\begin{array}{c} a_1^\T \\ a_2^\T \\ \vdots \\ a_n^\T \end{array}\rt)
= I_q.
\end{align}
This means that $V=\lt(\begin{array}{cccc} a_1 & a_2 & \cdots & a_n \end{array}\rt)^\T$ is an isometry. Then, since the column vectors of $V$ are orthonormal, we obtain
\begin{align}
U_1U_2\cdots U_mV = \lt(\begin{array}{cccc}
1 & 0 & \cdots & 0 \\
0 & 1 & \cdots & 0 \\
\vdots & \vdots & \ddots & \vdots \\
0 & 0 & \cdots & 1 \\
0 & 0 & \cdots & 0 \\
\vdots & \vdots & \vdots & \vdots \\
0 & 0 & \cdots & 0
\end{array}\rt) \in \mathbb{R}^{n\times q}
\end{align}
using $m=nq-\f12q(q+1)$ appropriate two-level orthogonal matrices $U_1,U_2,\hdots,U_m\in\mathbb{R}^{n\times n}$ (see Sec. 4.5.1 of \cite{nielsen10}). From this it follows that
\begin{align}
V = U_m^\T U_{m-1}^\T \cdots U_1^\T\lt(\begin{array}{cccc}
1 & 0 & \cdots & 0 \\
0 & 1 & \cdots & 0 \\
\vdots & \vdots & \ddots & \vdots \\
0 & 0 & \cdots & 1 \\
0 & 0 & \cdots & 0 \\
\vdots & \vdots & \vdots & \vdots \\
0 & 0 & \cdots & 0
\end{array}\rt) \in \mathbb{R}^{n\times q}.
\end{align}

Since each two-level orthogonal matrix can be specified with a single real parameter, the real rank-one measurement can be specified with $m=nq-\f12q(q+1)$ unconstrained real parameters. We can then use the algorithm for solving the unconstrained nonlinear programming problem to obtain the optimal measurement $M({}\cdot{};\theta)$ for the given $\theta$ and $G$ by using \eqref{eq:opt_m}.
For two point sources, we obtained the optimal measurements using the {\tt basinhopping} and {\tt Powell} algorithms of the {\tt SciPy} package.

Note that the parametrization for a rank-one measurement with a non-zero imaginary part can also be done in the same way \cite{yamagata_m}.

\begin{acknowledgments}
The first author is grateful to Professor Koichi Yamagata for providing him with Ref.~\cite{yamagata_m} and kindly allowing him to include portions of it in this paper.
This work was supported by JSPS KAKENHI Grant Numbers JP17H02861, JP20H02168 and JP22H00510.
\end{acknowledgments}


\bibliography{aps_submission}

\begin{thebibliography}{9}
\bibitem{note1} It is theoretically known that according to Fujiwara \cite{fujiwara06}, 16 values are sufficient, but in fact, according to Yamagata \cite{yamagata_m}, 12 values are sufficient.
\end{thebibliography}

\end{document}